\documentclass[preprint,prb,aps,floats,amssymb,showpacs,draft,%
floatfix,superscriptaddress]{revtex4} 
\usepackage{psfig}

\begin{document}

\title{ 
Static and dynamic structure factors \\ 
in three-dimensional randomly diluted Ising models
} 

\author{Pasquale Calabrese}
\affiliation{ 
Dipartimento di Fisica dell'Universit\`a di Pisa and INFN, \\
Largo Pontecorvo 2, I-56127 Pisa, Italy.  } 
\author{Andrea Pelissetto} 
\affiliation{Dipartimento di Fisica
  dell'Universit\`a di Roma ``La Sapienza" and INFN, 
  Piazzale Aldo Moro 2, I-00185 Roma, Italy}
\author{Ettore Vicari} 
\affiliation{ 
Dipartimento di Fisica dell'Universit\`a di Pisa and INFN, \\
Largo Pontecorvo 2, I-56127 Pisa, Italy.  }

\begin{abstract}
  
  We consider the three-dimensional randomly diluted Ising model and study the
  critical behavior of the static and dynamic spin-spin correlation functions
  (static and dynamic structure factors) at the paramagnetic-ferromagnetic
  transition in the high-temperature phase.
  We consider a purely relaxational dynamics without conservation
  laws, the so-called model A. 
  We present Monte Carlo simulations and perturbative field-theoretical
  calculations.  While the critical behavior of the static structure factor 
  is quite similar to that occurring in pure Ising systems, 
  the dynamic structure factor shows a substantially different critical 
  behavior. In particular, the dynamic correlation function shows a
  large-time decay rate which is momentum independent.  This effect
  is not related to the presence of the Griffiths tail, which is expected 
  to be irrelevant in the critical limit, but rather to the breaking of 
  translational invariance, which occurs for any sample and which,
  at the critical point, is not recovered even after the disorder average. 

\end{abstract}



\maketitle


\section{Introduction and Summary}
\label{intro}

The effect of disorder on magnetic systems remains, after decades of
investigation, a not fully understood subject.  It is then natural to
investigate relatively simple models, to try to understand the common
features of disordered systems.  In this regard, randomly
diluted spin systems are quite interesting. First, they represent 
simple models which describe the universal properties of the 
paramagnetic-ferromagnetic transition in uniaxial 
antiferromagnets with impurities \cite{Belanger-00}
and, in general, the order-disorder transition in 
Ising systems in the presence of uncorrelated local dilution. 
Second, they give the opportunity for investigating 
general problems concerning the effects of disorder on the 
critical behavior. 
Indeed, several important results, theoretical developments, and
approximation schemes found for these models have been later
generalized to more complex systems like spin glasses, quantum
disordered spin models, etc....  

In this paper we consider three-dimensional randomly diluted Ising (RDI)
systems. Their critical behavior has been extensively studied.
\cite{Belanger-00,PV-02,FHY-03} There is 
now ample evidence that the magnetic transition in these 
systems, if it is continuous, belongs to a unique universality class, and
several universal properties are now known quite accurately. 
Beside the static critical behavior, we also investigate the 
dynamic critical behavior, considering a 
purely relaxation dynamics without
conservation laws, the so-called model A,~\cite{HH-77} 
which is appropriate for uniaxial antiferromagnets. 
We focus on the dynamic (time-dependent) spin-spin correlation function
\begin{eqnarray}
G(x_2-x_1,t_2-t_1) \equiv 
\overline{\langle \sigma(x_1,t_1) \,\sigma(x_2,t_2) \rangle},
\label{twopi}
\end{eqnarray}
where $\sigma(x,t)$ is an Ising variable, the overline indicates the quenched
average over the disorder probability distribution, and $\langle
\cdot\cdot\cdot \rangle$ indicates the thermal average.  From the function
$G(x,t)$, one obtains the static (equal-time) structure factor
$\widetilde{G}(k)$ and the dynamic structure factor $\widehat{G}(k,\omega)$.
They are physically relevant quantities, which can be measured in neutron or
X-ray scattering experiments.~\cite{footBorn} It is therefore interesting to
study the effects of disorder on these physical quantities, and check whether
disorder gives rise to qualitative changes with respect to pure systems.  We
investigate their scaling behavior close to the magnetic transition for $T\to
T_c^+$ in the high-temperature phase. As we shall see, while the critical
behavior of the static structure factor is very similar to that in pure Ising
systems, the critical behavior of the dynamic structure factor is
significantly different; in particular, the large-momentum behavior shows some
new features.

Since the critical region in the paramagnetic phase, i.e. for $T\gtrsim T_c$,
is located in the Griffiths phase,\cite{Griffiths,v-06} it is mandatory to
discuss first the relevance of the so-called Griffiths singularities and
Griffiths tails for the universal critical behavior of the correlation
functions when $T\rightarrow T_c^+$.  In fact, one of the most notable
features of randomly diluted spin systems is the existence of the so-called
Griffiths phase for $T_c<T<T_p$, where $T_p$ is the critical temperature of
the pure system.  This is essentially related to the fact that, in the
presence of disorder, the critical temperature $T_c$ is lower than $T_p$, and
therefore, in the temperature interval $T_c<T<T_p$, there is a nonvanishing
probability to find compact clusters without vacancies (Griffiths islands)
that are fully magnetized.  They give rise to essential nonanalyticities in
thermodynamic quantities.\cite{Griffiths,v-06} Moreover, these clusters are
responsible for a nonexponential tail in dynamic correlation
functions.\cite{Bray-88,DRS-88,Bray-89,CMM-98} In the case of RDI systems one
can show that
\begin{equation}
G(x,t) \approx G_G(t) = B \exp [- C (\ln t)^{3/2} ], 
\label{G-non-exp}
\end{equation}
for any finite $x$ and $t\to \infty$, which implies a diverging relaxation
time.  We should mention that these effects are quite difficult to detect, and
there is still no consensus on their experimental evidence even in systems
with correlated disorder, in which these effects are magnified (see, e.g.,
Ref.~\onlinecite{expG,v-06} and references therein).

Griffiths essential singularities give quantitatively negligible effects on
thermodynamic quantities and on the static critical behavior.  One can argue
that also the Griffiths tail (\ref{G-non-exp}) is irrelevant in the critical
limit: the nonexponential tail does not contribute to the critical scaling
function associated with $G(x,t)$.\cite{Bray-88} This is essentially due to
the fact that $B$ and $C$ that appear in Eq.~(\ref{G-non-exp}) are expected to
be smooth functions of the temperature, approaching finite constants as $T\to
T_c$.  Thus, in the critical limit, $t\to\infty$, $T\to T_c$ at fixed
$t\xi^{-z}$, where $\xi\sim (T-T_c)^{-\nu}$ is the diverging correlation
length, the nonexponential contribution simply vanishes.  To understand why,
let us consider the simplified situation in which the contributions to the
autocorrelation function $G(x,t)$ due to the Griffiths islands and to the
critical modes just sum as
\begin{equation}
G(t)\approx G_C(t)+ G_G(t)=
a \xi^b \exp(- c t \xi^{-z}) + B \exp (- C (\ln t)^{3/2}),
\end{equation}
where we neglect all couplings between Griffiths and critical modes.
Here, $G_C(t)$ is the critical contribution, while $G_G(t)$
is the nonexponential Griffiths tail, which dominates for $t \gg t^*$, where
$t^*$ is the time at which the two terms have the same magnitude.  In the
critical limit we have $t^*\sim \xi^z (\ln \xi)^{3/2}$.  Since the critical
limit is taken at fixed $t/\xi^{z}$, the relevant quantity is $t^*/\xi^{z}$,
which diverges as $(\ln \xi)^{3/2}$ approaching the phase transition.  This
means that, for any fixed value of $t/\xi^z$,  the condition
$t\ll t^*$ is always satisfied
sufficiently close to the critical temperature $T_c$,
i.e.,  the nonexponential tail is negligible.

In order to determine the scaling behavior of the spin-spin correlation
function (\ref{twopi}), we perform Monte Carlo (MC) simulations of a
three-dimensional RDI model on a simple cubic lattice and perturbative
field-theoretical (FT) calculations. In the following we briefly summarize our
main results.

The high-temperature critical behavior of the static structure factor
$\widetilde{G}(k)$ is substantially analogous to that of $\widetilde{G}(k)$ in
pure Ising systems.\cite{MPV-02} We consider the universal scaling function
$g(Q^2)\equiv {\widetilde{G}(k)/\widetilde{G}(0)}$, where $Q^2\equiv k^2\xi^2$
and $\xi$ is the second-moment correlation length. We find that $g(Q^2)$ is
very well approximated by the Ornstein-Zernike form (Gaussian free-field
propagator) $g(Q)_{OZ}= 1/(1 + Q^2)$: deviations are less than 1\% for
$Q\lesssim 5$ and increase to 5\% at $Q \approx 50$.  At large momenta, for
$Q\gtrsim 10$ say, the static structure factor follows the Fisher-Langer
law;\cite{FL-68} in particular, $g(Q^2)\approx 0.92/Q^{2-\eta}$ with
$\eta\approx 0.036$ for $Q\gtrsim 30$.

At variance with the static case, the dynamic structure factor displays
substantial differences with respect to the pure case, even though, as
expected, the Griffiths tail turns out to be irrelevant in the critical limit.
We consider the universal scaling function
\begin{equation}
\Gamma(Q^2,S)\equiv \lim_{T\rightarrow T_c^+} 
{\widetilde{G}(k,t)\over \widetilde{G}(k,0)}
\label{scalfunc}
\end{equation}
in the critical limit $t\to\infty$, $k\to 0$, and $T\to T_c^+$ at fixed $Q$
and $S$.  Here $S\equiv t/\tau_{\rm int}$, where $\tau_{\rm int}$ is the
zero-momentum integrated autocorrelation time.  Perturbative field theory
shows that $\Gamma(Q^2,S)$ decays exponentially, as in the case of pure
systems, and this is confirmed by the simulation results.  In pure Ising
systems~\cite{CMPV-03-2} the large-$S$ decay rate of $\Gamma(Q^2,S)$ depends
on $Q$: the large-$S$ behavior is very similar to that of $\Gamma(Q^2,S)$ in
the noninteracting Gaussian model, i.e.  $\Gamma(Q^2,S) \sim \exp[-\kappa(Q^2)
S]$, where $\kappa(Q^2)=(1+Q^2)$.  This behavior drastically changes in the
presence of random impurities; in particular, the large-$S$ decay rate becomes
independent of $Q$.  MC simulations and FT perturbative calculations show
that, for generic values of $Q$, $\Gamma(Q^2,S)$ has two different behaviors
as a function of $S$.  For small values of $S$, it decreases rapidly, with a
rate that increases as $Q$ increases, as it does in the pure
system.\cite{CMPV-03-2} For large $S$ instead, $\Gamma(Q^2,S)$, and therefore
$\widetilde{G}(k,t)$, decreases with a momentum-independent rate. For large
$Q$ and $S$ we find
\begin{equation}
\Gamma(Q^2,S)\sim S^a Q^{-\zeta} e^{- \kappa S},
\label{largeqs}
\end{equation}
where $a$ and $\zeta$ are critical exponents, and $\kappa$ does not depend on
$Q$.  We present a physical argument which relates the different large-$S$
behavior compared to pure systems to the loss of translational invariance.
Such a phenomenon is obvious for a given fixed sample, but at the critical
point translational invariance is not even recovered after averaging over
disorder, because of the absence of self-averaging.  Note that this phenomenon
is only related to disorder and thus, it is expected in all systems in which
disorder is relevant.

In general, the perturbative calculations predict 
a scaling behavior of the form 
\begin{equation}
\Gamma(Q^2,S)\sim Q^{-\zeta} f_\zeta(S),
\label{Gamma-largeQ}
\end{equation}
for large $Q$,
where $f_\zeta(S)$ is a function of $S$ such that 
$f_\zeta(S) \sim S^a e^{-\kappa S}$ for $S\to \infty$.
This behavior implies that 
$G(x,t)$ is always nonanalytic for $x=0$ and any $t$. Indeed, 
because of Eq.~(\ref{Gamma-largeQ}), the integral
\begin{equation}
 \int d^d Q\, Q^{2n} \Gamma(Q^2,S) g(Q^2) \sim 
  \int d^d k\, k^{2n} \widetilde{G}(k,t)
\label{moments-Gamma}
\end{equation}
diverges for
$n \ge n_c\equiv (\zeta +2 - d -\eta)/2$ ($d$ is the
spatial dimension).  Since the moments of $\widetilde{G}(k,t)$ are directly
related to the derivatives of $G(x,t)$ with respect to $x$ computed for $x =
0$, the $n$-th derivative of $G(x,t)$
diverges for $n\ge n_c$; hence $G(x,t)$ is not analytic at $x=0$. This implies
that the scaling function $F(Y^2,S)$, defined as
\begin{equation}
G(x,t) = \xi^{-d + 2 - \eta} F(Y^2,S), \qquad Y^2 \equiv x^2/\xi^2,
\end{equation}
behaves as $F(Y^2,S) = f_0(S) + f_\lambda(S) |Y|^\lambda + \cdots$, where
$\lambda = \zeta + 2 - d - \eta$, for $Y^2\to 0$. MC simulations indicate that 
$\zeta\approx 2$ in three
dimensions, which implies $\lambda\approx 1$. 
This phenomenon does not occur in pure
systems, since in this case the decay rate $\kappa$ depends on $Q$ and
guarantees the integrability of the integrand which appears in
Eq.~(\ref{moments-Gamma}) for any $n$. Therefore, $F(Y^2,S)$ has an analytic
expansion around $Y^2=0$, i.e.,  $F(Y^2,S) = f_0(S) + f_2(S) Y^2 + \cdots$.  
This
nonanalyticity should be a general property of random models in which 
disorder is relevant, and not specific of RDI systems.

Finally, it is worth mentioning that in the case of three-dimensional randomly
dilute multicomponent spin models, such as the $XY$ and the Heisenberg model,
the effects of disorder found in RDI models are expected to be suppressed in
the critical limit $T\rightarrow T_c^+$, and should only appear as peculiar
scaling corrections.~\cite{footalpha} Indeed, the asymptotic critical behavior
of the correlation functions is expected to be the same as that in the
corresponding pure model, because the pure fixed point is stable under random
dilution (according to the Harris criterion,\cite{Harris-74} dilution is
irrelevant if the specific-heat exponent $\alpha$ of the pure system 
is negative).

The paper is organized as follows.  In Sec.~\ref{modelsec} we
introduce the model that we study.  In Sec.~\ref{sec.2} we discuss the static
structure factor in the high-temperature phase.  In Sec.~\ref{sec2.1} we
define the quantities that are computed in the MC simulation, in
Sec.~\ref{sec2.2} we present some perturbative calculations, while in
Sec.~\ref{sec2.3} we discuss the MC results.  In Sec.~\ref{sec.3} we discuss
the dynamic structure factor.  Again, we first define the basic quantities
(Sec.~\ref{sec3.1}), then we present a one-loop perturbative calculation
(Sec.~\ref{sec3.2}), and finally we report the MC results (Sec.~\ref{sec3.3}
and \ref{sec3.4}).  In the appendices we report some details of the
perturbative calculations.

\section{The model}
\label{modelsec}

We consider the randomly site-diluted Ising model with Hamiltonian
\begin{equation}
{\cal H}_{\rho} = - \sum_{<xy>}  \rho_x \,\rho_y \; \sigma_x \sigma_y,
\label{Hs}
\end{equation}
where the sum is extended over all pairs of nearest-neighbor sites of a simple
cubic lattice, $\sigma_x=\pm 1$ are Ising spin variables, and $\rho_x$
are uncorrelated quenched random variables, which are equal to 1
with probability $p$ (the spin concentration) and 0 with
probability $1-p$ (the impurity concentration).  
For $p_s < p < 1$, where $p_s$ is the site-percolation point
($p_{s}=0.3116081(13)$ on a simple cubic lattice~\cite{BFMMPR-99}),
the model has a continuous transition with a ferromagnetic low-temperature
phase.

MC simulations of RDI systems have shown rather conclusively (see, e.g.,
Refs.~\onlinecite{Belanger-00,PV-02,FHY-03,BFMMPR-98,HPPV-07,HPV-07})
that their continuous transitions belong to a single
universality class.  The RDI universality
class has been extensively studied by using 
FT methods and MC simulations. 
At present, the most accurate estimates of
the critical exponents are \cite{HPPV-07} $\nu = 0.683(2)$ and $\eta =
0.036(1)$, obtained by a finite-size analysis of MC data.
These estimates are in good agreement with those obtained by using
field theory. An analysis of the six-loop perturbative expansions in
the three-dimensional massive zero-momentum scheme gives \cite{PV-00}
$\nu=0.678(10)$ and $\eta=0.030(3)$.  Note the good agreement between
FT and MC results, in spite of the fact that the perturbative FT series
for dilute systems are not Borel
summable.~\cite{BMMRY-87,McKane-94,AMR-00} Also the
correction-to-scaling exponents have been determined quite accurately.
For the leading exponent $\omega$, MC simulations give \cite{HPV-07}
$\omega = 0.29(2)$ (older simulations gave $\omega =
0.33(3)$\cite{HPPV-07} and $\omega = 0.37(6)$\cite{BFMMPR-98}), while
field theory predicts $\omega = 0.25(10)$,\cite{PV-00} $\omega =
0.32(6)$.\cite{PS-00} For the next-to-leading exponent $\omega_2$, an
appropriate analysis of the FT expansions gives\cite{HPPV-07} $\omega_2
= 0.82(8)$, which is consistent with the MC results for improved
models.\cite{CMPV-03,HPPV-07} Beside the critical exponents, also the
equation of state,\cite{CPV-eqst} some amplitude
ratios,~\cite{CPV-eqst,CPV-cross,BCBJ-04} the universal crossover
functions between the pure and the RDI fixed point 
\cite{CPV-cross} and between the Gaussian and the RDI fixed
point,\cite{CPV-cross,FHY-00} and the crossover exponent 
in the presence of a weak random magnetic field\cite{cpv-03} 
have been computed.

A full characterization of the types of disorder that lead to a
transition in the RDI universality class is still lacking. For
instance, RDI transitions also occur in systems that are not
ferromagnetic: this is the case of the Edwards-Anderson model ($\pm J$
Ising model), which is frustrated for any amount of
disorder. \cite{EA-75} Nonetheless, the paramagnetic-ferromagnetic
transition line that starts at the pure Ising transition point and
ends at the multicritical Nishimori point belongs to the RDI
universality class.\cite{HPPV-07-2}

Beside the static behavior, we also consider the critical behavior of
a purely relaxation dynamics without conservation laws, the so-called
model A, as appropriate for uniaxial magnets.\cite{HH-77} The critical
behavior of the model-A dynamics for RDI systems has been recently
studied numerically (for a critical review of the existing results,
see Ref.~\onlinecite{HPV-07}). It has been shown that the relaxational dynamics
belongs to a single dynamic universality class,
\cite{HPV-07} characterized by the dynamic critical exponent $z=2.35(2)$.

For Hamiltonian (\ref{Hs}) an accurate study of the dependence of the size
of the corrections to scaling on $p$ is reported in Ref.~\onlinecite{HPPV-07}.
It turns out that the leading scaling corrections associated with
$\omega=0.29(2)$ are suppressed for $p = p^* = 0.800(5)$, in agreement with
the findings of Refs.~\onlinecite{BFMMPR-98,CMPV-03}.  For this reason we have
performed our simulations at $p = 0.8$.

\section{Static structure factor in the high-temperature phase}
\label{sec.2}

\subsection{Definitions} \label{sec2.1}

We consider the static (equal-time) two-point correlation function
$G(x) \equiv G(x,t=0)$.
In the infinite-volume limit
we define the {\em second-moment} correlation length $\xi$ 
\begin{equation}
\xi^2 \equiv 
    - {1\over \chi} \left. {\partial \widetilde{G}(k) \over \partial k^2} 
      \right|_{k^2 = 0} , 
\label{xidefc}
\end{equation}
where $\widetilde{G}(k)$ is the Fourier transform of $G(x)$ and
\begin{equation}
\chi\equiv\sum_x G(x) = \widetilde{G}(0)
\label{chidef}
\end{equation}
is the magnetic susceptibility.  
It is also possible to define an {\em exponential} correlation length
$\xi_{\rm exp}$.
Given the infinite-volume $G(x)$, we define
\begin{equation}
\xi_{\rm exp} \equiv  - \lim_{|x|\to\infty} {|x|\over \ln G(x)}.
\end{equation}
In the critical limit $\xi$ and $\xi_{\rm exp}$ diverge.
If $t_r \equiv (T - T_c)/T_c$ and $T_c$ is the critical temperature,
for $|t_r|\to 0$ we have in the thermodynamic limit
\begin{equation}
\xi, \xi_{\rm exp}\sim |t_r|^{-\nu},
\label{diver-s}
\end{equation}
where $\nu$ is a  universal critical exponent. In the same limit, 
correlation functions have a universal behavior. For instance,
the infinite-volume 
$\widetilde{G}(k)/\chi$ becomes a universal function of the scaling variable
\begin{equation}
Q^2 \equiv k^2 \xi^2, \qquad\qquad 
\end{equation}
i.e. we can write in the scaling limit $k \to 0$, $t_r \to 0$ at fixed $Q$
\begin{equation}
{\chi^{-1} \widetilde{G}(k)} \approx  g(Q^2),
\label{scaling-statico}
\end{equation}
where $g(x)$ is universal. Moreover, the ratio $\xi^2/\xi_{\rm exp}^2$ 
converges to a universal constant $S_M$ defined by
\begin{equation}
\xi^2/\xi_{\rm exp}^2 \approx S_M.
\end{equation}
For a Gaussian theory the spin-spin correlation function shows the
Ornstein-Zernike (OZ) behavior
\begin{equation}
\widetilde{G}_{OZ}(k) = {Z\over k^2 + r}.
\end{equation}
It follows $\chi = Z/r$, $\xi^2 = 1/r$, and
\begin{equation}
g_{OZ}(Q^2) = {1 \over 1 + Q^2}\; .
\label{OZ}
\end{equation}
Moreover, $S_M = 1$.

Fluctuations change this behavior. For small $Q^2$, $g(Q^2)$ is analytic,
so that we can write the expansion
\begin{equation} 
g(Q^2)^{-1} = 1 + Q^2 + \sum_{n=2} c_n Q^{2n},
\label{gQ-small}
\end{equation}
where the coefficients $c_n$ parametrize the deviations from the OZ behavior.
For large $Q^2$, the structure factor behaves as 
\begin{equation}
g(Q^2) \approx {C_1\over Q^{2 - \eta}}
  \left(1 + {C_2\over Q^{(1-\alpha)/\nu}} +
            {C_3\over Q^{1/\nu}} + \cdots \right),
\label{eq:FL}
\end{equation}
a behavior predicted theoretically by Fisher and Langer 
\cite{FL-68} and proved in the FT
framework in Refs.~\onlinecite{BAZ-74,BLZ-74}. 

\subsection{Field-theory results} \label{sec2.2}

We determined the coefficients $c_n$ by using two different
FT approaches: the $\sqrt{\epsilon}$-expansion
approach, in which the renormalization-group 
parameters are computed as series in powers
of $\sqrt{\epsilon}$, $\epsilon = 4-d$, and the massive zero-momentum (MZM) 
approach, in which one works directly in three dimensions.
We computed the first few coefficients $c_n$ to
$O(\epsilon^{3/2})$ in the $\sqrt{\epsilon}$ expansion, and to four loops in
the MZM scheme.  The corresponding expansions are reported in App.~\ref{AppA}. 
Setting $\epsilon = 1$ 
in the $\sqrt{\epsilon}$ expansions (\ref{exp-cn-sqrte}), we obtain
$c_2=-4\times 10^{-4}$, $c_3=1.0\times 10^{-5}$, $c_4=-4\times 10^{-7}$, 
and $c_5=2\times 10^{-8}$. 
In the MZM approach, resumming the perturbative 
expansions (\ref{cnMZM}) as discussed in 
Ref.~\onlinecite{PV-00}, we obtain
\begin{eqnarray}
&&c_2=-4(1)\times 10^{-4}, \label{c2-est}\\
&&c_3=1.2(3)\times 10^{-5}, \\
&&c_4=-5(2)\times 10^{-7}.
\end{eqnarray}
These results are fully consistent with those obtained in the
$\sqrt{\epsilon}$ expansion, in spite of the fact that in that case we
have not applied any resummation and we have simply set $\epsilon =
1$. We can also compute $S_M$. Since the coefficients $c_n$ are very
small, we obtain
\begin{equation}
S_M \approx 1 + c_2 = 0.9996(1),
\label{SM-est}
\end{equation}
where we used the estimate (\ref{c2-est}) of $c_2$.
As in the Ising case,\cite{CPRV-98,MPV-02} the coefficients $c_n$ 
show the pattern
\begin{equation}
|c_n|\ll |c_{n-1}|\ll...\ll |c_2| \ll 1\qquad\qquad 
{\rm for}\qquad n\geq 3.
\label{patternci}
\end{equation}
This is consistent with the expected analyticity properties of 
$\widetilde{G}(k)$. 
Since the complex-plane singularity
in $\widetilde{G}(k)^{-1}$ that is closest to the origin is expected to be the 
three-particle cut located at $k =  \pm 3i/\xi_{\rm exp}$,\cite{FS-75,Bray-76}
the function $g(Q^2)^{-1}$ is analytic up to $Q^2 = - 9 S_M$. It follows that 
$c_n \approx - c_{n-1}/(3 \sqrt{S_M})$, at least asymptotically.

The large-$Q$ behavior can be investigated in the $\sqrt{\epsilon}$ expansion.
The three-loop calculation of the two-point function
reported in App.~\ref{appsqe} allows us to determine the 
perturbative expansion of the coefficients $C_i$ appearing in 
Eq.~(\ref{eq:FL}).
Setting $\epsilon = 1$ in the expressions (\ref{lqc}), 
we obtain $C_1 \approx 0.95$, $C_2 + C_3 = -0.96$.

In order to compare with the experimental and numerical data it is important
to determine $g(Q^2)$ for all values of $Q$.
For the pure Ising structure factor, several interpolations have been proposed
with the correct large- and small-$Q$ behavior.
\cite{FB-67,TF-75,FS-75,Bray-76,FB-79,BBC-82,MPV-02}
The most successful one is due to Bray,\cite{Bray-76}
which incorporates the expected singularity structure of $g(Q^2)$.
In this approach, one
assumes $1/g(Q^2)$ to be well-defined in the complex
$Q^2$ plane, with a cut on the negative real $Q^2$ axis,
starting at the three-particle cut $Q^2 = - r^2$ with
$r^2 = 9 S_M$. Then, one obtains the spectral representation
\begin{eqnarray}
H(Q^2) &\equiv & \int_{r}^\infty du\, u^{1-\eta} {F(u) \over Q^2 + u^2} ,
\nonumber \\
{1\over g(Q^2)} &=& 1 + {Q^2 H(Q^2)\over S_M H(-S_M)},
\label{gQ-dispersive}
\end{eqnarray}
where $F(u)$ is the spectral function, which must satisfy
$F(+\infty) = 1$, $F(u) = 0$ for $u< r$, and $F(u)\ge 0$ for
$u\ge r$.

In order to obtain an approximation one must specify $F(u)$.
Bray \cite{Bray-76} proposed to use a spectral function that gives
exactly the Fisher-Langer asymptotic behavior, i.e.
\begin{equation}
F_{B}(u) = {P_1(u) - P_2(u) \cot {1\over2} \pi \eta \over
          P_1(u)^2 + P_2(u)^2},
\end{equation}
where
\begin{eqnarray}
P_1(u) &=& 1 + {C_2\over u^p} \cos {\pi p\over 2} +
             {C_3\over u^{1/\nu}} \cos {\pi\over 2\nu},
\nonumber \\
P_2(u) &=&  {C_2\over u^p} \sin {\pi p\over 2} +
             {C_3\over u^{1/\nu}} \sin {\pi\over 2\nu},
\label{PQ-def}
\end{eqnarray}
with $p\equiv (1-\alpha)/\nu$. To obtain a numerical expression we fix 
$\nu = 0.683$, $\eta = 0.036$,\cite{HPPV-07} and use the 
estimate (\ref{SM-est}) of $S_M$.
We must also fix $C_2$ and $C_3$. Bray proposes to fix
$C_2+C_3$ to its $\epsilon$-expansion value (in our case $C_2 + C_3 = -0.96$)
and then to determine these constants by requiring $F_B(u=r) = 0$.
These conditions give $C_2 = -8.04$ and $C_3 = 7.07$.
As a check, we can compare the estimates of
$c_n$ and $C_1$ obtained by using Bray's approximation $g_B(Q^2)$
with the previously quoted results.
We obtain
\begin{equation}
C_1 = {2 \sin \pi\eta/2\over \pi} S_M H(-S_M) \approx 0.92,
\label{BrayC1}
\end{equation}
and $c_2 \approx -4 \cdot 10^{-4}$, $c_3 \approx 9\cdot 10^{-6}$, 
$c_4 \approx -4\cdot 10^{-7}$.
These results are in very good agreement with 
those obtained before.

\subsection{Monte Carlo results}  \label{sec2.3}

In this section we study Hamiltonian (\ref{Hs}) at $p=0.8$
in the high-temperature phase, with the purpose of determining
the infinite-volume spin-spin correlation function $\widetilde{G}(k)$. We 
perform simulations on lattices of size $32\le L\le 256$ in the range 
$0.275\le \beta \le 0.2856$ [note that\cite{HPV-07} $\beta_c = 0.2857431(3)$]. 
The number of samples varies with $N$, being of the order 
of 3000, 10000, 30000, 40000 for $L=256,128,64,32$. For each sample, we start 
from a random configuration, run 1000 Swendsen-Wang and 1000
Metropolis iterations for thermalization, and then 
perform 2000 Swendsen-Wang sweeps. At each iteration
we measure the correlation function 
$G(x;\beta,L)$ and the structure factor $\widetilde{G}(k;\beta,L)$.
Since rotational invariance is recovered in the 
critical limit, to speed up the Fourier transforms, we determine it as 
\begin{equation}
\widetilde{G}(k;\beta,L) = {1\over 3} \sum_{x,y,z} 
   (e^{ikx} + e^{iky} + e^{ikz}) 
   \overline{\langle \sigma(0,0,0) \sigma(x,y,z) \rangle},
\end{equation}
where the sum runs over the coordinates $(x,y,z)$ of the lattice sites.
Of course, on a finite lattice $k$ can only assume the values 
$2\pi n/L$, where $n$ is an integer such that $0\le n \le L - 1$.
We also compute the second-moment correlation length $\xi(\beta,L)$
defined by
\begin{equation}
\xi(\beta,L)^2 \equiv {\widetilde{G}(0;\beta,L) - 
          \widetilde{G}(k_{\rm min};\beta,L) \over 
          \hat{k}_{\rm min}^2 \widetilde{G}(k_{\rm min};\beta,L) },
\label{xidef}
\end{equation}
where $k_{\rm min} \equiv 2\pi/L$, $\hat{k} \equiv 2 \sin k/2$.
For $L\to \infty$, $\xi(\beta,L)$ converges to the infinite-volume definition
(\ref{xidefc}) with $L^{-2}$ corrections.

In order to determine $g(Q^2)$ we go through several different steps. 
First, for each $\beta$ and $L$ we interpolate the numerical data 
in order to obtain $\widetilde{G}(k;\beta,L)$ for any $k$ in the 
range $[0,\pi]$. For this purpose we fit the numerical results for 
$h(k;\beta,L) \equiv  \widetilde{G}(0;\beta,L)/\widetilde{G}(k;\beta,L)$ to
\begin{equation}
h(k;\beta,L) = 1 + \sum_{n=1}^{n_{\rm max}} a_n \hat{k}^{2n}, 
\qquad \hat{k} = 2 \sin {k\over2}.
\end{equation}
We increase $n_{\rm max}$ until the sum of the residuals ($\chi^2$) is 
less than half of the fitted points (those corresponding to $1\le n \le L-1$), 
i.e. $\chi^2 < L/2$ (note that the 
data are strongly correlated and thus it makes no sense to require 
$\chi^2/{\rm DOF}\approx 1$, where ${\rm DOF} = L-1-n_{\rm max}$
is the number of degrees of freedom of the fit).
In most of the cases we take $n_{\rm max} = 5$, but in a few cases we had to 
take $n_{\rm max}$ as large as 10.

\begin{figure*}[tb]
\begin{tabular}{cc}
{\psfig{width=8truecm,angle=-90,file=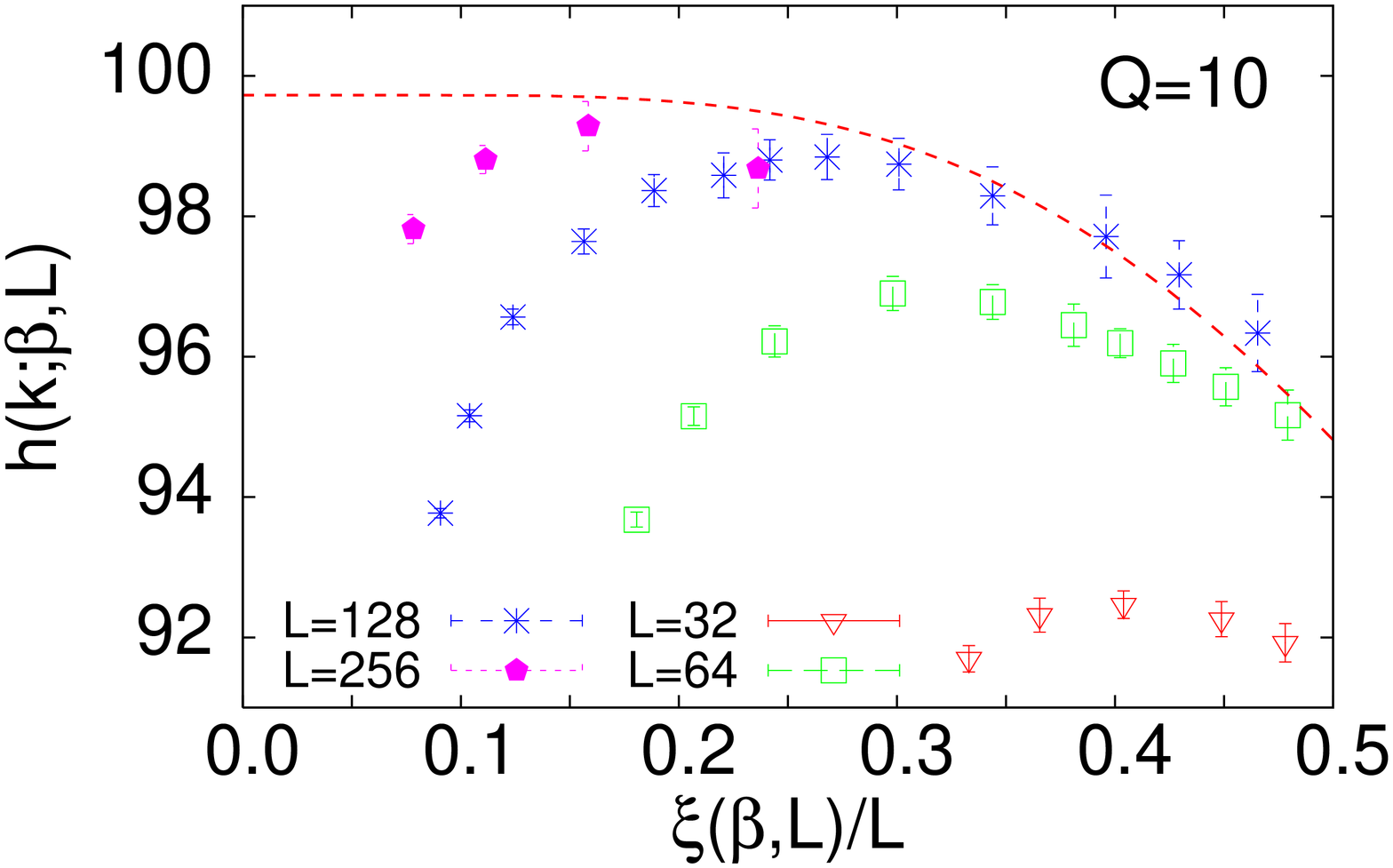}} &
{\psfig{width=8truecm,angle=-90,file=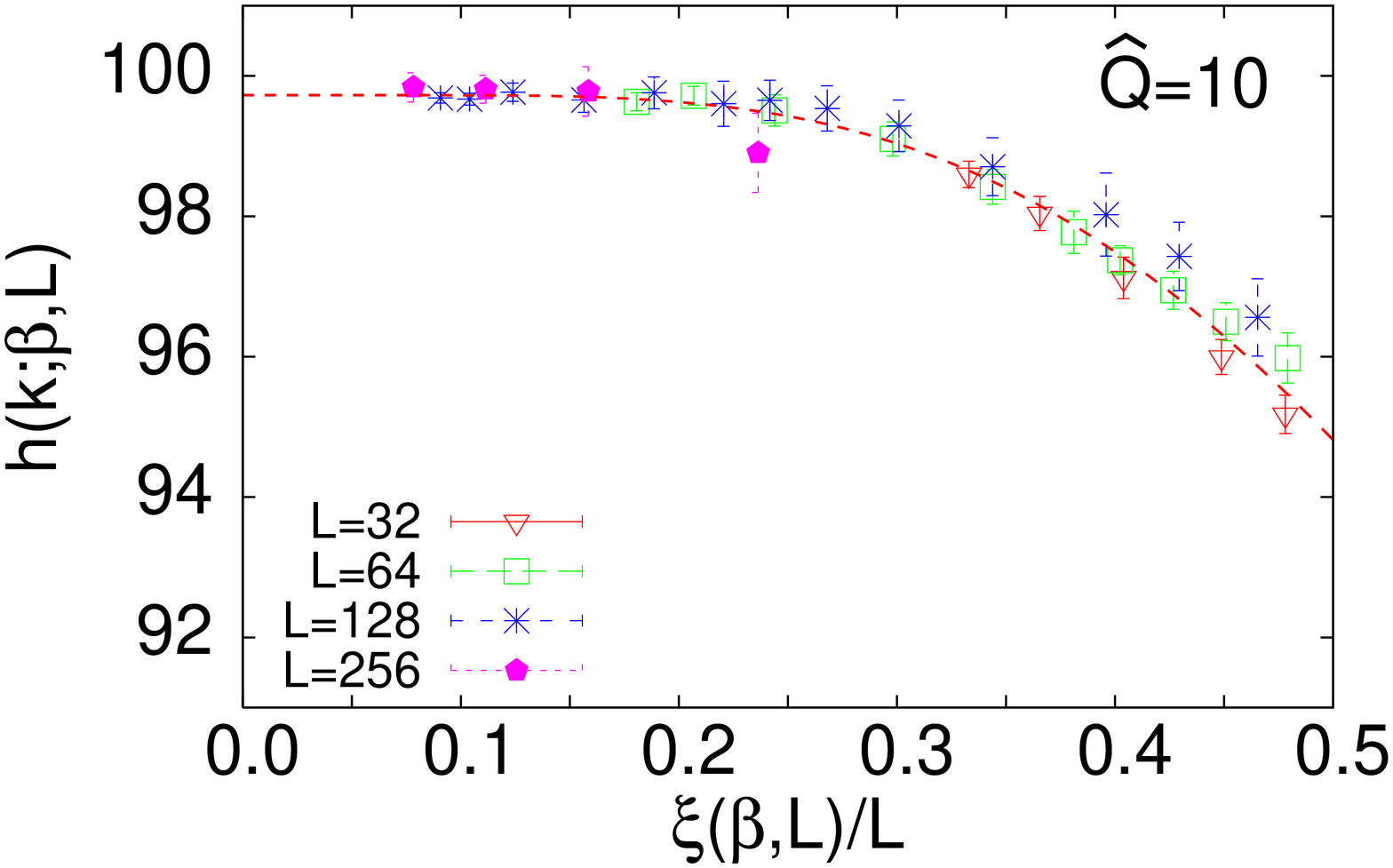}} \\
\end{tabular}
\vspace{2mm}
\caption{(Color online)
Estimates of $h(k;\beta,L)$ vs $\xi(\beta,L)/L$ at fixed 
$Q = 10$ (left) and $\widehat{Q} = 10$ (right). 
We only report data corresponding to $k \le k_{\rm max} = \pi/3$. 
We also report an interpolating curve (dashed line), 
which is obtained by fitting all data 
at fixed $\widehat{Q}$ reported on the right, as explained in the text. 
}
\label{fig:Fqx-comp}
\end{figure*}

Then, we investigate the finite-size effects. In the critical limit
we expect
\begin{equation}
h(k;\beta,L) \approx 
    F\left(Q \equiv k \xi(\beta,L),{\xi(\beta,L)\over L}\right).
\label{h-FSS1}
\end{equation}
Equivalently, one can also use 
\begin{equation}
h(k;\beta,L) \approx 
 F\left(\widehat{Q} \equiv \hat{k} \xi(\beta,L),{\xi(\beta,L)\over L}\right),
\label{h-FSS}
\end{equation}
where $\hat{k} = 2\sin k/2$. The two scaling forms are equivalent in 
the scaling limit $k\to 0$, $L\to \infty$, $\xi(\beta,L)\to \infty$ 
at fixed $Q$ (or $\widehat{Q}$) and $\xi(\beta,L)/L$; as a consequence,
the function $F(x,y)$ is the same in the two cases. 
Indeed, $\hat{k} = k + O(k^3)$, and thus, by keeping fixed $\widehat{Q}$ 
or $Q$, one only changes analytic corrections decaying as $L^{-2}$. 
In particular, whatever choice is made, the structure factor $g(Q^2)$
is equal to $1/F(Q,0)$. Apparently, the corrections
we are talking about here are less relevant than the nonanalytic 
corrections that should decay as $L^{-\omega_2}$, $\omega_2 = 0.82(8)$, 
and thus, {\em a priori} one would expect only small differences between the 
two approaches. Instead, as we show below, only by keeping $\widehat{Q}$ fixed 
is one able to determine the structure factor in the infinite-volume limit.

In Fig.~\ref{fig:Fqx-comp} we show the numerical data for $Q = 10$ (left)
and $\widehat{Q} = 10$ (right). On the left one observes very large size 
corrections which make impossible in practice the determination of the 
infinite-volume limit $\xi(\beta,L)/L \to 0$. On the right instead, 
there are no significant 
scaling corrections and all data fall approximately on a single curve. 
Size corrections are small for $\xi(\beta,L)/L \lesssim 0.20$ and 
the extrapolation to $\xi(\beta,L)/L \to 0$ is feasible. 
In the two panels 
we also show the interpolation of the data at fixed $\widehat{Q}$.
As expected, the data at fixed $Q$ converge to this interpolation, but it is 
clear that no real information could have been obtained on the 
infinite-volume limit from the data in the left panel. In order to clarify 
why scaling at fixed $\widehat{Q}$ is so much better than scaling at 
fixed $Q$, we consider the lattice Gaussian model with nearest-neighbor 
couplings. In this case, the spin-spin correlation function on a finite 
lattice is given by
\begin{equation}
\widetilde{G}_G(k) = {Z\over \hat{k}^2 + r},
\end{equation}
so that $\xi^2 = 1/r$ and 
\begin{equation}
h(k) = {1 + \widehat{Q}^2}.
\end{equation}
Thus, if we take the finite-size scaling 
limit at fixed $\widehat{Q}$ there are no finite-size 
corrections: the scaling is exact on any finite lattice. On the other hand,
at fixed $Q$ we obtain
\begin{equation}
h(k) \approx (1 + Q^2) \left[
   1 - {1\over 12 L^2} {Q^4 (L^2/\xi^2)\over 1 + Q^2} + \cdots\right].
\end{equation}
In this case we have $1/L^2$ corrections, which diverge as 
$\xi/L\to 0$, exactly as we observe in our data. These corrections
moreover increase with $Q$  and thus make it difficult, if not 
impossible, to estimate the structure factor. 

\begin{figure*}[tb]
\begin{tabular}{c}
{\psfig{width=9truecm,angle=-90,file=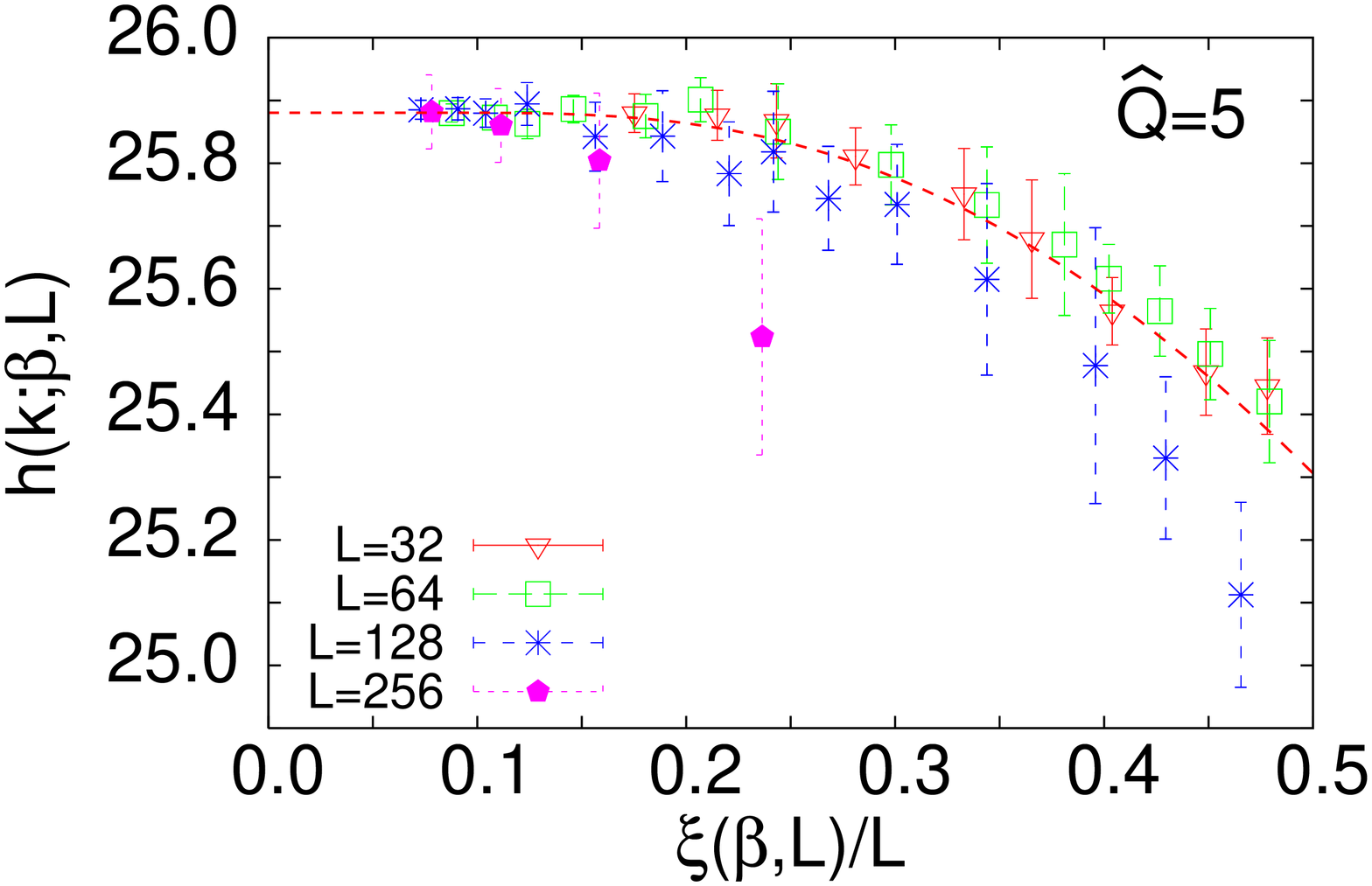}} \\
{\psfig{width=9truecm,angle=-90,file=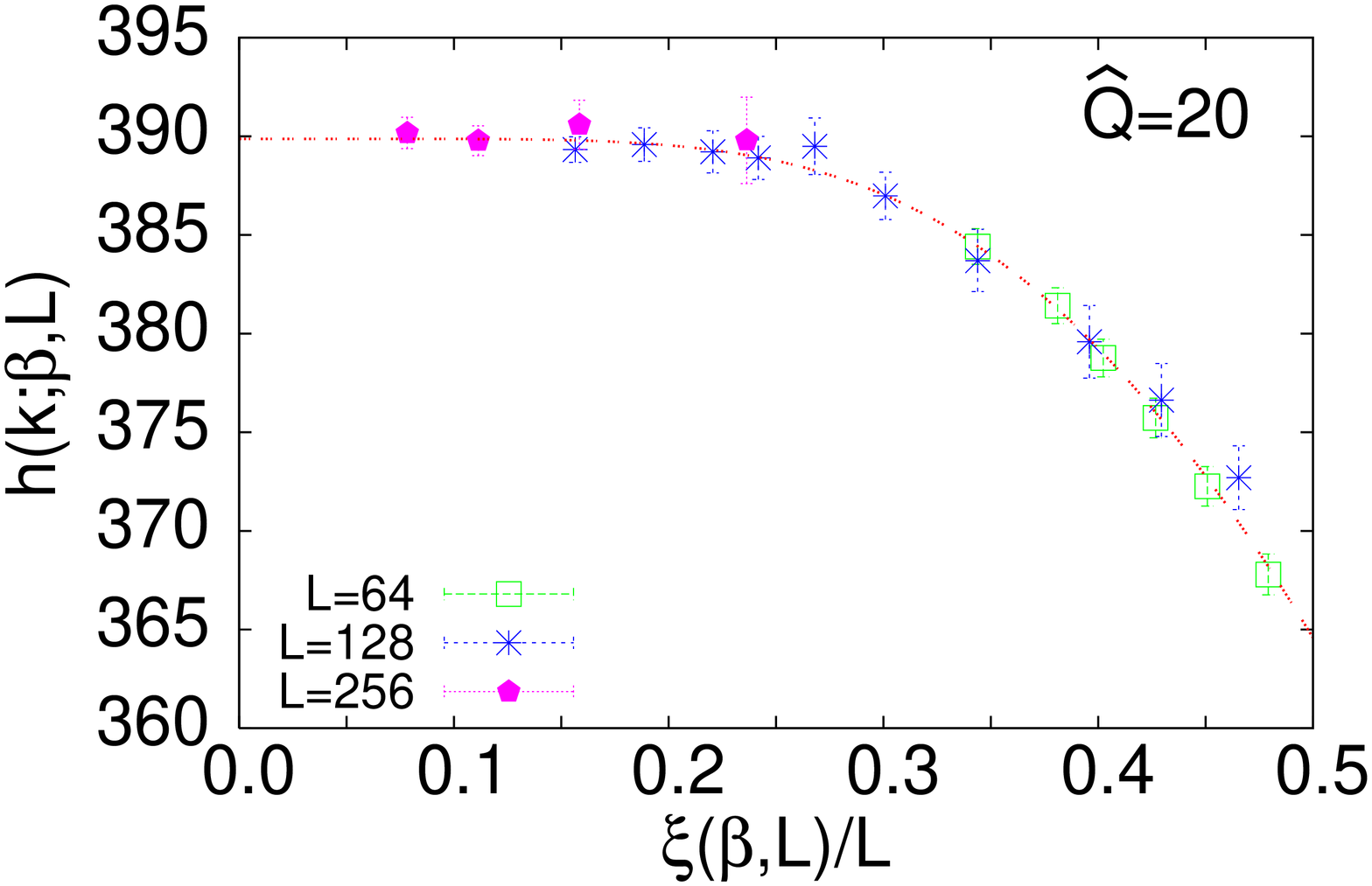}} \\
{\psfig{width=9truecm,angle=-90,file=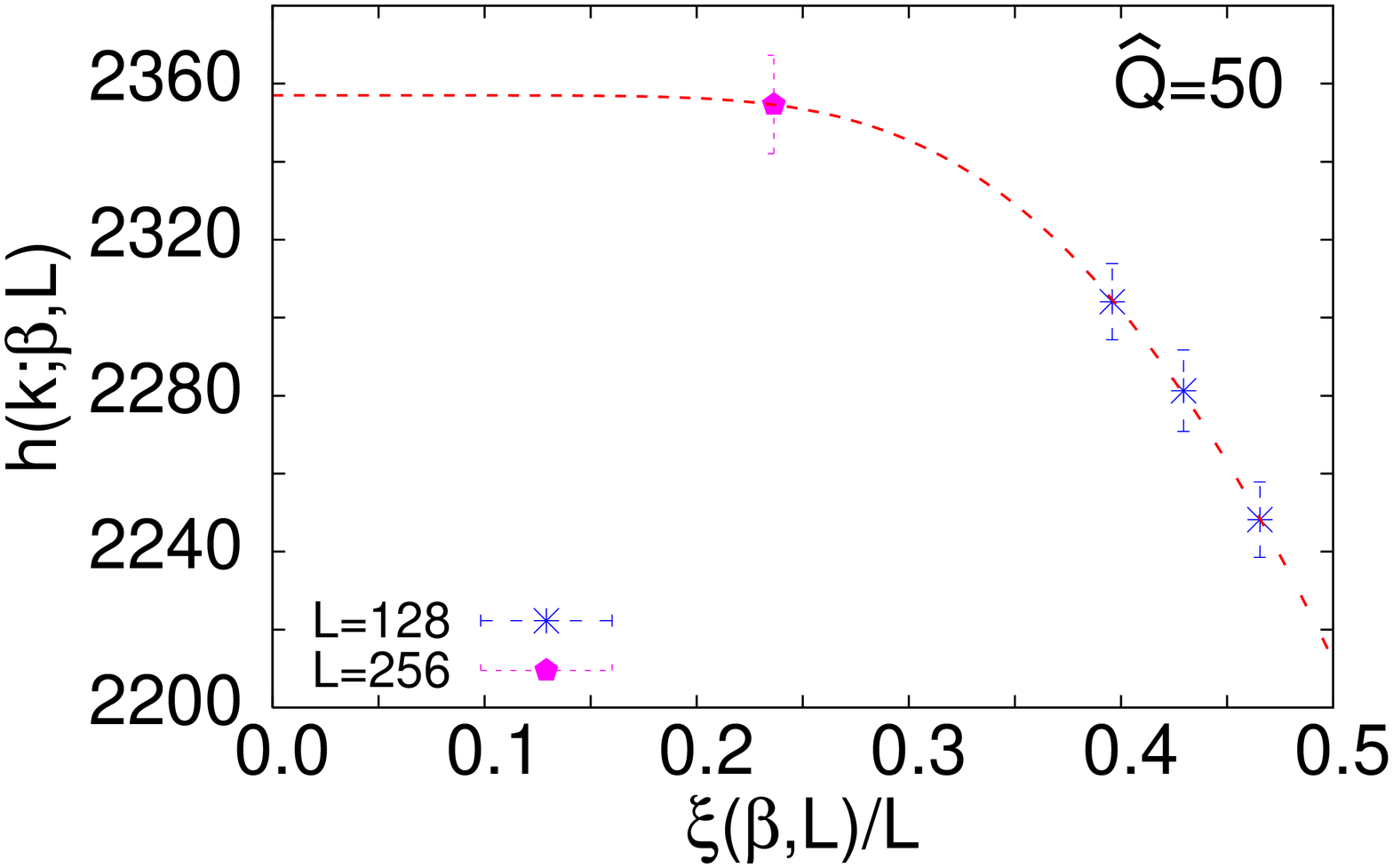}} \\
\end{tabular}
\vspace{2mm}
\caption{(Color online)
MC results for $h(k;\beta,L)$ vs $\xi(\beta,L)/L$ at fixed 
$\widehat{Q}$ for three different values of $\widehat{Q}$:
5 (top), 20 (middle), 50 (bottom). Only data satisfying 
$k \le k_{\rm max} = \pi/3$ are reported. 
The interpolation (dashed line) corresponds to a fit 
of the data with $L\ge 64$ as described in the text.
}
\label{fig:Fqx-variq}
\end{figure*}

As a consequence of the above-reported discussion we consider below 
the finite-size scaling limit at fixed $\widehat{Q}$. 
In Fig.~\ref{fig:Fqx-variq} we show $h(k;\beta,L)$ for $\widehat{Q} = 5,20,50$.
Since $\hat{k}$ can be at most 2, for each $\widehat{Q}$ we can only consider 
values of $\beta$ and $L$ such that $\xi(\beta,L) \ge \widehat{Q}/2$. 
However, since the 
critical limit is obtained for $k\to 0$, results close to the 
antiferromagnetic point $k = \pi$ cannot have a good scaling behavior. 
Therefore, in the analysis we have only considered values of $k$ such that 
$k\le k_{\rm max}$. If $k_{\rm max}$ varies between $\pi/4$ and $\pi/3$,
the final results are essentially
independent of $k_{\rm max}$. The data reported in Figs.~\ref{fig:Fqx-comp} 
and \ref{fig:Fqx-variq}
scale as predicted by Eq.~(\ref{h-FSS}). Within the precision of our results
some corrections  to scaling are only visible for $\widehat{Q} = 5$ and 
$\xi(\beta,L)/L\gtrsim 0.2$. They however die out fast in the interesting limit
$\xi(\beta,L)/L \to 0$. Note also that, as $\widehat{Q}$ increases, 
the number of available points decreases and indeed we are not able to go 
beyond $\widehat{Q}\approx 50$ with our data.

In order to determine the infinite-volume limit $F(Q,0)$, 
we have taken all 
data satisfying $\xi(\beta,L)/L \le 0.5$ and we have fitted them to
\begin{equation}
\left. h(k;\beta,L) \right|_{\hat{k} \xi = \widehat{Q}}  = 
   a_0 + \sum_{j=1}^{j_{\rm max}} a_j \exp[-jL/\xi(\beta,L)].
\label{fitting-FSS}
\end{equation}
The fitting form (\ref{fitting-FSS}) is motivated by theory, which 
predicts exponentially small finite-size corrections in the 
high-temperature phase. With the precision of our data it is 
sufficient to take $j_{\rm max} = 2$ to obtain $\chi^2/{\rm DOF}\lesssim 1$. 
The coefficient $a_0$ allows us to estimate $g(Q^2)$:
$g(Q^2) = 1/F(Q,0) = 1/a_0$. 
The results for $k_{\rm max} = \pi/4$ and $\pi/3$ are essentially identical
within errors up to $Q\approx 40$ (for $k_{\rm max} = \pi/4$
we do not have enough data to determine reliably $F(Q,0)$ for  
$Q\gtrsim 40$). 
In the following we take those corresponding to $k_{\rm max} = \pi/3$,
which allow us to compute $g(Q^2)$ up to $\widehat{Q} = 50$. In order to 
detect scaling corrections we have repeated the analysis including each time
only data such that $L\ge L_{\rm min}$. The results are essentially
independent of $L_{\rm min}$. For instance, for $\widehat{Q} = 5$, one of the 
values we considered in Fig.~\ref{fig:Fqx-variq} (in this case some scaling
corrections are present for $\xi(\beta,L)/L\gtrsim 0.20$), we obtain 
$a_0 = 25.881(6)$, 25.881(7), 25.882(9) for 
$L_{\rm min} = 32, 64, 128$, respectively. This is due to the fact 
that $a_0$ is determined by the results at small values of 
$\xi(\beta,L)/L$ and in this range there are essentially no scaling 
corrections. In the following we choose conservatively $L_{\rm min} = 64$.

\begin{figure*}[tb]
\centerline{\psfig{width=10truecm,angle=-90,file=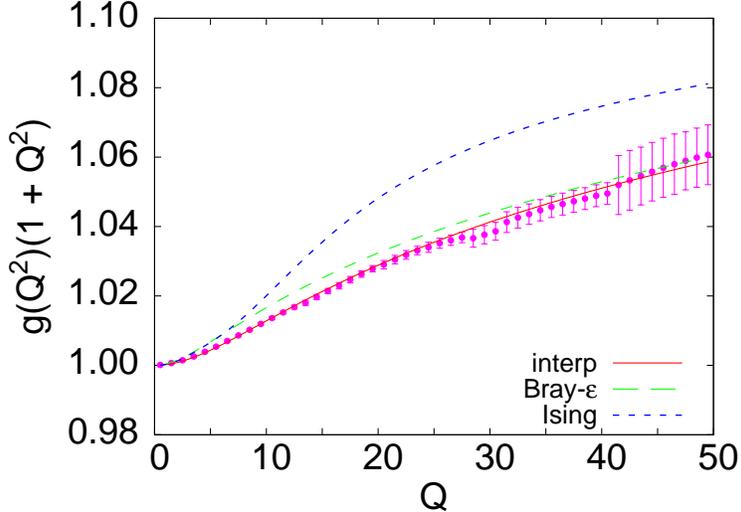}}
\vspace{2mm}
\caption{(Color online)
Estimates of the scaling function $g(Q^2)(1 + Q^2)$ for integer values of $Q$. 
We also report Bray's approximation, 
in which $C_2 + C_3$ is fixed to the $\sqrt{\epsilon}$ value 
(Bray-$\epsilon$),
and the structure factor 
for the pure Ising model (Ising). The curve "interp" (solid line)
corresponds to the 
interpolation $g_{\rm int}(Q^2)$ reported in Eq.~(\ref{GQ2-interp}).
}
\label{fig:gQ2}
\end{figure*}

Our final estimate of $g(Q^2)$ is reported in Fig.~\ref{fig:gQ2}. 
Deviations from the OZ behavior are quantitatively small and indeed at $Q = 50$ 
the relative deviation is only 0.05. 
It is important to note that the estimates of $g(Q^2)$ 
at different values of $Q$ are 
correlated since the estimates of $\widetilde{G}(k;\beta,L)$ for 
different values of $k$ are statistically correlated. This explains the 
regularity of the results. Note also that the error changes rather abruptly 
in a few cases. For instance, this occurs between $Q = 40$ and $Q = 41$. 
This happens because at $Q = 40$, the estimate of $a_0$ is essentially 
determined
by the result obtained for $\beta = 0.2853$, $L = 256$, which corresponds 
to $\xi(\beta,L)/L = 0.158$ and $\hat{k} = 0.989$. For $Q = 41$ this lattice is 
no longer considered, since the corresponding 
$k$ exceeds $k_{\rm max} = \pi/3$ ($\hat{k}_{\rm max} = 1$). 
For $Q = 41$, the result with the smallest $\xi(\beta,L)/L$ corresponds to 
$\xi(\beta,L)/L = 0.236$. The extrapolation to the infinite-volume limit 
is therefore much more imprecise.

For $Q\to \infty$, $g(Q^2)\approx C_1/Q^{2-\eta}$, see Eq.~(\ref{eq:FL}). We
fit the estimates of $\ln g(Q^2)$ reported in Fig.~\ref{fig:gQ2} (they
correspond to integer values of $Q$ between 1 and 50) to $a + (\eta - 2)\ln
Q$. If we include only data with $Q > Q_{\rm min} = 15$ and 20, we obtain
$\eta = 0.032(1)$, 0.032(2), respectively.  The error we quote here assumes
that all data are independent, which is not the case. In order to determine
the correct error bar, one should take into account the covariance among the
results at different values of $Q$.  This is not easy and therefore, in order
to estimate the role of the statistical correlations, we use a more
phenomenological approach.  If $g_{\rm est}(Q^2)$ is the estimate of $g(Q^2)$
and $\sigma(Q^2)$ the corresponding error, we consider new data $g_{\rm
  est}(Q^2) - \sigma(Q^2)$ with the same error and we repeat the fit. We
obtain $\eta = 0.029$ and $\eta = 0.027$ for $Q_{\rm min} = 15$ and 20.
Analogously, if we consider $g_{\rm est}(Q^2) + \sigma(Q^2)$, we obtain $\eta
= 0.035, 0.037$.  This simple analysis indicates that $\pm0.005$ is a
plausible estimate of the statistical error. Therefore, we quote $\eta =
0.032(5)$ as our final result.  This estimate is in good agreement with
that reported in Ref.~\onlinecite{HPPV-07}, $\eta = 0.036(1)$, obtained from a
finite-size scaling analysis of the susceptibility.  In order to estimate
$C_1$, we consider $g(Q^2) Q^{2-\eta}$, fixing $\eta$ to $\eta = 0.036(1)$.
\cite{HPPV-07}
For $Q \gtrsim 20$ this quantity is essentially constant: $g(Q^2) Q^{2-\eta}
= 0.921(1)$, 0.920(1), 0.917(2), 0.919(3), for $Q = 20,25,30,35$. We thus take
\begin{equation}
   C_1 = 0.919(3)[3]
\label{est-C1-MC}
\end{equation}
as our final estimate.  The error in brackets gives the variation of the 
estimate as $\eta$ varies by one error bar ($\pm 0.001$).
This estimate is close to the FT result $C_1 \approx 0.95$ and in perfect 
agreement with the estimate (\ref{BrayC1}) 
obtained by using Bray's approximation for the
spectral function, $C_1 \approx 0.92$. Indeed, as can be seen in 
Fig.~\ref{fig:gQ2}, Bray's interpolation
represents a very good approximation of the 
numerical data, deviations being quite tiny. 

In Fig.~\ref{fig:gQ2} we also report the structure factor in pure 
Ising systems (we use the phenomenological approximation reported in 
Ref.~\onlinecite{MPV-02}, see their Eq.~(30) with $Q_{\rm max} = 15$ and 
$n_{\rm max} = 6$). In the pure case, deviations from the OZ behavior 
are larger: the addition of impurities has the effect of reducing the 
deviations from the OZ behavior.

Finally, we report a phenomenological interpolation which reproduces 
well our numerical data and is consistent with the large $Q^2$ behavior, 
$g(Q^2)\approx 0.919 Q^{0.036}/(1 + Q^2)$:
\begin{equation}
g_{\rm int}(Q^2) = 
   {( 1 + 0.0227953 Q^2  + 0.0000839355 Q^4 )^{0.009} \over 
    1 + Q^2}.
\label{GQ2-interp}
\end{equation}

\section{Dynamic structure factor in the high-temperature phase}
\label{sec.3}

In this section we consider the dynamic behavior of the Metropolis
algorithm, which is 
a particular example of a relaxational dynamics without conservation laws,
the so-called model A, as appropriate for 
magnetic systems. In Ref.~\onlinecite{HPV-07} we computed the dynamic critical
exponent, obtaining $z=2.35(2)$. Here, we focus on the dynamic structure
factor.

\subsection{Definitions} \label{sec3.1}

To investigate the dynamic behavior we consider the 
time-dependent two-point correlation function (\ref{twopi})
and its Fourier transform $\widetilde{G}(k,t)$ with respect to the $x$
variable.  
Then, we define the integrated autocorrelation time
\begin{equation}
\tau_{\rm int}(k) \equiv {1\over2} 
 \sum_{t=-\infty}^\infty \, {\widetilde{G}(k,t)\over \widetilde{G}(k,0)} = 
  {1\over 2} + \sum_{t=1}^\infty \, 
                 {\widetilde{G}(k,t)\over \widetilde{G}(k,0)},
\label{tauintdef}
\end{equation}
and the exponential autocorrelation time
\begin{equation}
\tau_{\rm exp}(k) \equiv - \lim_{|t|\to\infty}
      {|t|\over \ln \widetilde{G}(k,t)},
\label{tauedef}
\end{equation}
which controls the large-$t$ behavior of $\widetilde{G}(k,t)$.
Here $t$ is the Metropolis time and one time unit corresponds to a complete 
lattice sweep.

Beside $\tau_{\rm int}(k)$ and $\tau_{\rm exp}(k)$ we also define 
autocorrelation times $\tau_{{\rm int},x}$ and 
$\tau_{{\rm exp},x}$.\cite{HPV-07} In general, given an autocorrelation 
function $A(t)$ we define 
\begin{eqnarray}
&&I(s) \equiv  {1\over 2} + {1\over A(0)} \sum_{t=1}^s A(t), \\
&& \tau_{\rm eff}(s) \equiv {n\over \ln [A(s-n/2)/A(s+n/2)]},
\label{defBL}
\end{eqnarray}
for any integer $s$ and any fixed even $n$. By linear interpolation these
functions can be extended to any real $s$. 
Then, we define $\tau_{{\rm int},x}$ and
$\tau_{{\rm exp},x}$ as the solutions of the consistency equations
\begin{eqnarray}
&& \tau_{{\rm exp},x} = \tau_{\rm eff}(x\tau_{{\rm exp},x}),
\label{tauxexp-def} 
\\
&& \tau_{{\rm int},x} = I(x\tau_{{\rm int},x}).
\label{tauxint-def}
\end{eqnarray}
These definitions have been discussed in
Ref.~\onlinecite{HPV-07}. There, it was shown that they provide
effective autocorrelation times with the correct critical
behavior. For $x\to\infty$, $\tau_{{\rm exp},x}$ and $\tau_{{\rm
int},x}$ converge to $\tau_{{\rm exp}}$ and $\tau_{{\rm int}}$,
respectively.

As discussed in the introduction, for $T_c < T \le T_p$ the
correlation function $G(x,t)$ does not decay exponentially for any
finite value of $x$, but presents a slowly decaying tail, 
cf.~Eq.~(\ref{G-non-exp}). Therefore, $\tau_{\rm exp}(k)$ diverges for
all $T_c \le T < T_p$. As discussed in
Ref.~\onlinecite{HPV-07}, this is not the case for the effective
exponential autocorrelation time $\tau_{{\rm exp},x}$, which is finite
for any finite $x$.  Note that correlation functions decaying as in
Eq.~(\ref{G-non-exp}) have a finite time integral and thus the
integrated autocorrelation time is finite.

In the critical limit the autocorrelation times diverge.  If $t_r
\equiv (T - T_c)/T_c$ and $T_c$ is the critical temperature, for
$|t_r|\to 0$ we have
\begin{equation}
\tau_{\rm int}(k) \sim \tau_{{\rm exp},x}(k) 
     \sim \tau_{{\rm int},x}(k) \sim |t_r|^{-z\nu} \sim \xi^z,
\label{diver}
\end{equation}
where $\nu$ is the usual static exponent and $z$ is a dynamic exponent
that depends on the considered dynamics: $\nu = 0.683(2)$ and $z =
2.35(2)$ in the present case.\cite{HPPV-07,HPV-07} In the same limit,
$\widetilde{G}(k,t)/\widetilde{G}(k,0)$ becomes a universal function
of the scaling variables
\begin{equation}
Q^2 \equiv k^2 \xi^2, \qquad\qquad 
S \equiv   t/\tau_{\rm int}(0),
\end{equation}
i.e. we can write 
\begin{equation}
{\widetilde{G}(k,t)\over \widetilde{G}(k,0)} = \Gamma(Q^2,S),
\end{equation}
where $\Gamma(Q^2,S)$ is universal, even in $S$, i.e., 
$\Gamma(Q^2,S) = \Gamma(Q^2,-S)$,
and satisfies the normalization conditions
\begin{equation}
\Gamma(Q^2,0) = 1, \qquad \qquad
\int_{0}^\infty \Gamma(0,S) dS = 1.
\label{norm-Gamma0}
\end{equation}
The function $\widetilde{G}(k,0)$ is the static structure factor whose
critical behavior has been discussed in Sec.~\ref{sec2.1}. Using
Eq.~(\ref{scaling-statico}) we can write $\widetilde{G}(k,t) = \chi
g(Q^2) \Gamma(Q^2,S)$. Analogously, we have
\begin{equation}
{\tau_{\rm int}(k)\over \tau_{\rm int}(0)} \equiv f_{\rm int}(Q^2),\qquad\qquad
\end{equation}
where the scaling function $f_{\rm int}(Q^2)$ is universal and
satisfies $f_{\rm int}(0) = 1$.

It is important to note that Eq.~(\ref{G-non-exp}) does not
necessarily imply that the scaling function $\Gamma(Q^2,S)$ decays
nonexponentially. On the contrary, as argued in Sec.~\ref{intro}, the
Griffiths tail (\ref{G-non-exp}) becomes irrelevant in the critical
limit. In view of that discussion it is natural to define a scaling function 
\begin{equation}
f_{\rm exp}(Q^2) \equiv  - \lim_{|S|\to\infty} 
          {|S|\over \ln \Gamma(Q^2,S)},
\label{fexp-Q2-def}
\end{equation}
which we call, rather loosely, the scaling function associated with 
the exponential autocorrelation time. Indeed, if 
$f_{\rm exp}(Q^2)$ is finite, for $S\to\infty$ we have
\begin{equation}
\Gamma(Q^2,S) \sim S^a \exp(-S/f_{\rm exp}(Q^2)),
\end{equation}
where $a$ is some critical exponent. In terms of quantities that are 
directly accessible numerically, we can define it as
\begin{equation} 
f_{\rm exp}(Q^2) = \lim_{x\to\infty} \lim_{k\to0;\xi\to\infty}
         {\tau_{{\rm exp},x}(k)\over \tau_{\rm int}(0)}.
\end{equation}
Of course, the two limits cannot be interchanged.

The dynamic structure factor $\widehat{G}(k,\omega)$ is defined as 
\begin{equation}
\widehat{G}(k,\omega) = \int_{-\infty}^\infty dt\,
            \widetilde{G}(k,t) e^{i \omega t} =
            2 \int_{0}^\infty dt\, \widetilde{G}(k,t) \cos \omega t.
\end{equation}
In the scaling limit we introduce a new scaling function 
$\sigma(Q^2,w)$ defined by 
\begin{eqnarray}
\sigma(Q^2,w) \equiv {\widehat{G}(k,\omega) \over 
\tau_{\rm int}(0) \widetilde{G}(k,0) } \qquad\qquad
 w \equiv \omega \tau_{\rm int}(0).
\label{defsigma}
\end{eqnarray}
The function $\sigma(Q^2,w)$ is essentially the ratio of the dynamic and 
static structure factors and is directly related to $\Gamma(Q^2,S)$:
\begin{equation}
\sigma(Q^2,w) = 2 \int_0^\infty dS\, \Gamma(Q^2,S)\, \cos wS.
\end{equation}
It is even in $w$ and satisfies the normalization conditions:
\begin{equation}
\sigma(0,0) = 2, \qquad\qquad 
\int_{-\infty}^\infty {dw\over 2\pi}\, \sigma(Q^2,w) = 1.
\end{equation}
Moreover, we have $\sigma(Q^2,0) = 2 f_{\rm int}(Q^2)$.

For a Gaussian theory the spin-spin correlation function is given by
\begin{equation}
\widetilde{G}_G(k,t) = {Z e^{-\Omega (k^2 + r) |t|}\over k^2 + r}.
\label{Gkt-G}
\end{equation}
It follows $\tau_{\rm int}(k) = [\Omega (k^2 + r)]^{-1}$, so that 
\begin{equation}
\Gamma(Q^2,S) = e^{-(Q^2 + 1) |S|}, \qquad
f_{\rm int}(Q^2) = f_{\rm exp}(Q^2) = {1\over 1 + Q^2}.
\label{Gamma-Gaussiano}
\end{equation}
Finally, we have
\begin{equation}
\sigma(Q^2,w) = {2(1 + Q^2) \over w^2 + (1 + Q^2)^2}.
\end{equation}

\subsection{Field-theory results} \label{sec3.2}

The dynamic structure factor can be computed in perturbation theory. 
The explicit one-loop calculation is reported in App.~\ref{AppB}. 
Two facts should be noted. First, perturbation theory predicts 
an exponential decay for $\Gamma(Q^2,S)$ for any $Q^2$. This is consistent
with the argument presented in the introduction, which predicted the 
absence of the Griffiths tail in the critical scaling functions. 
Second, one-loop 
perturbation theory predicts $f_{\rm exp}(Q^2)$ to be independent of $Q^2$. 
We wish now to argue that this result is exact and is related
to the breaking of translational invariance in disordered systems. Indeed,
consider the spin-spin correlation function for a given disorder 
configuration $\{\rho\}$,
\begin{equation}
  \gamma(x_1,x_2;t_1 - t_2;\{\rho\}) \equiv
    \langle \sigma(x_1,t_1) \sigma(x_2,t_2)\rangle_\rho,
\end{equation}
and the corresponding Fourier transform
\begin{equation}
  \widetilde{\gamma}(k_1,k_2;t_1 - t_2;\{\rho\}) =
   \sum_{x_1x_2} e^{ik_1x_1 + i k_2 x_2} \gamma(x_1,x_2;t_1 - t_2;\{\rho\}).
\end{equation}
In pure systems translational invariance implies that 
$\widetilde{\gamma}(k_1,k_2;t_1 - t_2;\{\rho\})$ vanishes unless 
$k_1 = - k_2$. This is not the case in disordered systems, where translational
invariance is lost. The average of $\widetilde{\gamma}$ over disorder
vanishes for $k_1\not= - k_2$ 
[it indeed corresponds to $\widetilde{G}(k,t_1-t_2)$], and thus 
translational invariance is somewhat recovered. However, this does not mean
that the critical theory is translationally invariant. For instance,
consider
\begin{equation}
\overline{\hphantom{?} 
   |\widetilde{\gamma}(k_1,k_2;t_1 - t_2;\{\rho\})|^2\hphantom{?}}.
\label{trans-rotta}
\end{equation}
It can be easily verified in perturbation theory that this quantity 
is not zero for {\em any} $k_1$ and $k_2$. Note that this breaking 
of translational invariance survives in the infinite-volume limit only close
to the critical point. In the paramagnetic phase, far from the critical
transition, self-averaging occurs and thus also the quantity 
(\ref{trans-rotta}) vanishes for $k_1\not= - k_2$ when $L\to\infty$.

Let us now show that, if translational invariance (both for the 
Hamiltonian and the transition rates) holds, the decay rate is
$k$ dependent: modes corresponding to different momenta decouple. 
Indeed, following Refs.~\onlinecite{Abe-68,DRS-88}, 
let ${\cal L}$ be the Liouville operator associated with the dynamics,
and $\lambda_a$ and $\psi_a$ be the corresponding 
eigenvalues and eigenvectors. Then, we have the spectral representation
($t > 0$)
\begin{equation}
 \widetilde{G}(k,t) = \sum_a e^{-\lambda_a t} 
   |\langle \sigma(k) | \psi_a \rangle|^2,
\end{equation}
where the sum runs over all eigenstates with nonvanishing eigenvalue of 
${\cal L}$. Here we have introduced the inner product 
\begin{equation}
  \langle f | g \rangle = \sum_\alpha \pi_\alpha f_\alpha^* g_\alpha,
\end{equation}
where $f$ and $g$ are functions defined over the configuration space, 
$\pi_\alpha$ is the equilibrium distribution, and the sum runs over 
all configurations $\alpha$ of the system. If the system
is translationally invariant, ${\cal L}$ commutes with the 
generator $T$ of the translations; hence, the eigenstates of 
${\cal L}$ are also eigenstates of $T$. Thus, we have 
decoupled sectors corresponding to different values of the momentum $k$
and therefore we have
\begin{equation}
 \widetilde{G}(k,t) = \sum_a e^{-\lambda_a(k) t} 
   |\langle \sigma(k) | \psi_a(k) \rangle|^2,
\end{equation}
where the sum runs over the eigenstates of momentum $k$. Hence,
if $\lambda_1(k)$ is the smallest eigenvalue in each sector, we have 
$\widetilde{G}(k,t)\sim e^{-\kappa t}$ with
$\kappa = \lambda_1(k)$; hence, the decay rate is $k$ dependent. 
If translational invariance is lost, all eigenfunctions contribute to
each single value of $k$. Note, however, that this does not 
necessarily imply that the decay rate 
$\kappa$ in Eq.~(\ref{largeqs}) is $Q$ independent. 
Indeed, one should average 
over the disorder distribution and this average could wash out the 
effect. We expect this to happen in the infinite-volume limit 
at fixed $T$, for $T\not=T_c$.
The perturbative results show that this is not the case at the 
critical point. Hence, all modes are coupled in the critical limit 
and $\kappa$ is momentum independent. This argument indicates that 
the $Q$-independence of $\kappa$ is strictly related to the breaking of 
self-averaging at the critical point and thus, we expect a similar phenomenon 
to occur for the low-temperature critical dynamical structure factor.

\begin{figure*}[tb]
\centerline{\psfig{width=9truecm,angle=-90,file=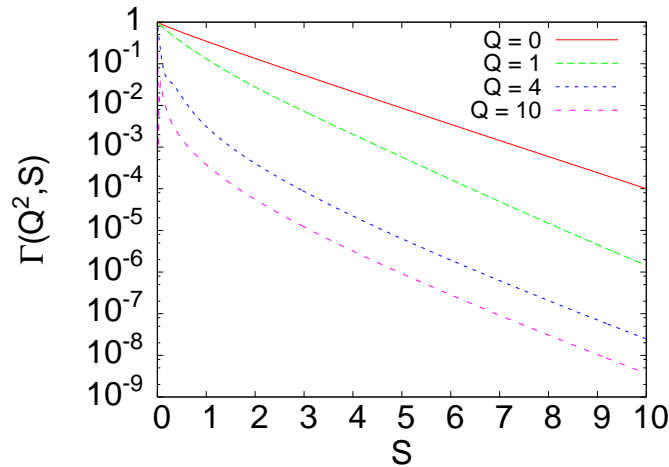}}
\vspace{2mm}
\caption{(Color online)
Scaling function $\Gamma(Q^2,S)$ as a function of 
$S$, as obtained in one-loop perturbation theory. 
}
\label{fig:Gkt-pert}
\end{figure*}

In Fig.~\ref{fig:Gkt-pert} we report 
$\Gamma(Q^2,S)$ as obtained by using 
Eqs.~(\ref{Gscal-HT}) and (\ref{tauexp-tauint}) 
and simply setting $\epsilon = 1$. 
The behavior we observe is quite different from what 
is observed in the Gaussian model. In this case,
Eq.~(\ref{Gamma-Gaussiano}) implies
$\ln \Gamma(Q^2,S) = - (1 + Q^2) |S|$. As a consequence, 
with a logarithmic vertical scale,
the data fall on straight lines with increasing slope as $Q^2\to\infty$.
Here instead, $\Gamma(Q^2,S)$ first decreases rapidly and then bends so that 
the large-$S$ decay is $Q^2$-independent. This behavior
is also very different from
that observed in the pure Ising model, whose dynamical critical behavior is 
very close to that of the Gaussian model.\cite{CMPV-03-2}

If $f_{\rm exp}(Q^2) = f_{\rm exp}$ is independent of $Q^2$,
for $S\to\infty$  we expect a behavior of the form
\begin{equation}
\Gamma(Q^2,S) \approx f(Q^2) S^a \exp(-S/f_{\rm exp}),
\label{Gamma-largeS}
\end{equation}
where $a$ is a critical exponent. At one loop, the calculations reported in 
App.~\ref{AppB} give $a = 0$ for $Q^2=0$,
$a= - 1$ for $Q^2\not=0$, and $f(Q^2) \sim Q^{-2}$ for $Q\to\infty$.
In general, we expect $f(Q^2)$ to vanish with a nontrivial exponent 
in the large-$Q$ limit and thus
we write
\begin{equation}
f(Q^2) \sim Q^{-\zeta},
\label{f-largeQ-main}
\end{equation}
with a new exponent $\zeta$.

Given $\widetilde{G}(k,t)$, one can compute $G(x,t)$, which can be 
written in the scaling form
\begin{equation}
G(x,t) = \xi^{-d + 2 - \eta} F(Y^2,S), \qquad Y^2 \equiv x^2/\xi^2.
\label{fys}
\end{equation}
Perturbation theory, see App.~\ref{AppB}, indicates that 
$F(Y^2,S)$ is not analytic for $Y^2\to 0$. It predicts a behavior
of the form
\begin{equation}
   F(Y^2,S) = f_0(S) + f_\lambda(S) |Y|^\lambda + \cdots,
\label{FYT-smallY-main}
\end{equation}
where $\lambda$ is a new exponent that can be related to 
the exponent $\zeta$ which appears in Eq.~(\ref{f-largeQ-main}):
$\lambda = \zeta - 1-\eta$ (in $d$ dimensions, as we discuss in App.~\ref{AppB},
$\lambda = \zeta + 2 - d-\eta$).
The exponent $\lambda$ is positive (hence $\zeta$ must be larger 
than $1+\eta$), since 
$G(x=0,t)$ is always finite. The quantity $f_\lambda(S) |Y|^\lambda$ 
represents a subleading nonanalytic correction to the leading term $f_0(S)$. 

\begin{figure*}[tb]
\centerline{\psfig{width=9truecm,angle=-90,file=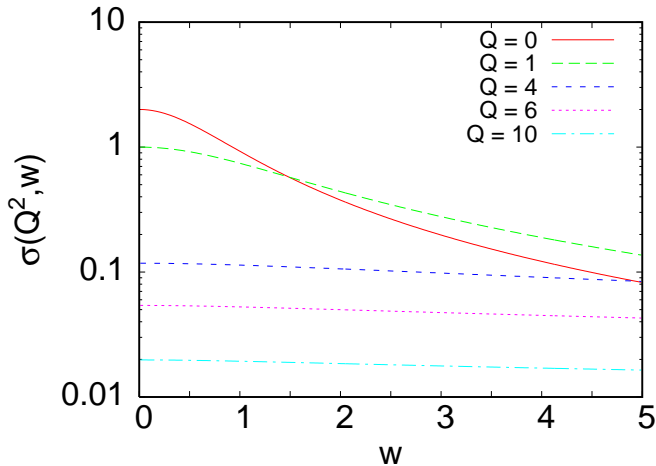}}
\vspace{2mm}
\caption{(Color online)
Scaling function $\sigma(Q^2,w)$ as a function of 
$w \equiv \omega\tau_{\rm int}(0)$, as obtained
in one-loop perturbation theory.
}
\label{fig:Gom-pert}
\end{figure*}

Finally, in Fig.~\ref{fig:Gom-pert} we report the 
one-loop perturbative expression of $\sigma(Q^2,w)$.
Note that the width of $\sigma(Q^2,w)$ does not decrease with
increasing $Q^2$, as it does in the Gaussian model.  
This is a consequence of the 
large-$S$ behavior of $\Gamma(Q^2,S)$, whose decay
is independent of $Q^2$.

\subsection{Simulation details} \label{sec3.3}

In this section we study the critical dynamics of Hamiltonian (\ref{Hs}) 
at $p=0.8$
in the high-temperature phase, with the purpose of determining
the time-dependent spin-spin correlation function $\widetilde{G}(k,t)$ and the 
related dynamic structure factor $\widehat{G}(k,\omega)$. We 
perform simulations on lattices of size $L\le 128$ in the range 
$0.275\le \beta \le 0.284$, corresponding to 
$4\lesssim \xi \lesssim 16$. For each disorder sample, we start
from a random configuration, run 1000 Swendsen-Wang and 1000
Metropolis iterations for thermalization, and then 
$N_{\rm it}$ Metropolis sweeps [typically, we took
$N_{\rm it}$ varying between $30 \tau_{\rm int}(0)$ and 
$100 \tau_{\rm int}(0)$]. The number of 
samples varies between 5000 and 20000. We measure the second-moment
correlation length $\xi(\beta,L)$ defined in Eq.~(\ref{xidef}) and 
the correlation function $\widetilde{G}(k,t)$. 
As we did for the static structure factor, 
we determine $\widetilde{G}(k,t)$ as 
\begin{equation}
\widetilde{G}(k,t) = {1\over 3} \sum_{x,y,z} 
   (e^{ikx} + e^{iky} + e^{ikz}) 
   \overline{\langle \sigma(0,0,0;0) \sigma(x,y,z;t) \rangle},
\end{equation}
where the sum runs over the coordinates $(x,y,z)$ of the lattice sites;
the time $t$ is expressed in units of Metropolis lattice sweeps.

Given $\widetilde{G}(k,t)$, we determine $\tau_{\rm int}(0)$. 
More precisely, we determine $\tau_{{\rm int},x}(0)$ with $x=5$, as 
defined by the self-consistent equation (\ref{tauxint-def}). 
As discussed above, this is a good autocorrelation
time for any $x$; therefore, we use this quantity to obtain a 
high-temperature estimate of $z$. We have also determined 
$\tau_{{\rm int},x}(0)$ with $x=8$. The results for $x=5$ and $x=8$ are 
consistent within errors, indicating that we can take 
$\tau_{{\rm int},5}(0)$ as an estimate of $\tau_{\rm int}(0)$.
We also consider the effective exponents $\tau_{{\rm exp},x}(0)$ defined 
by Eqs.~(\ref{defBL}) and (\ref{tauxexp-def}) with 
$A(t) = \widetilde{G}(k=0,t)$.
The results we quote correspond to $n=2$. 

\begin{table}
\squeezetable
\caption{MC results for the randomly site-diluted 
Ising model at $p=0.8$. We report the 
number of samples $N_s$, the number of Metropolis iterations in 
equilibrium $N_{\rm it}$, the second-moment correlation length $\xi$, the
zero-momentum integrated autocorrelation time 
$\tau_{\rm int}(0)$, and the effective 
zero-momentum exponential autocorrelation times $\tau_{{\rm exp},x}(0)$,
for $x=1,2$.
}
\label{tab:tabHT}
\begin{ruledtabular}
\begin{tabular}{crrrllll}
$\beta$ & $L$ & 
\multicolumn{1}{c}{$N_s$} & $N_{\rm it}$ & 
\multicolumn{1}{c}{$\xi$} & 
\multicolumn{1}{c}{$\tau_{\rm int}(0)$} &
\multicolumn{1}{c}{$\tau_{{\rm exp},1}(0)$} &
\multicolumn{1}{c}{$\tau_{{\rm exp},2}(0)$} \\
\hline
0.275 & 32 & $20000$ & $30000$ & 4.452(4)  & 36.66(27)  &  37.92(18) & 39.9(5)\\
0.278 & 32 & $20000$ &  $5000$ & 5.601(4)  & 62.25(23)  &  65.70(25) & 71.3(6)\\
      & 64 & $20000$ &  $5000$ & 5.622(2)  & 61.98(21)  &  65.41(30) & 69.4(9)\\
0.280 & 32 & $20000$ &  $8000$ & 6.872(7)  & 102.1(6)   &  106.8(5)&118.2(1.2)\\
0.281 & 32 & $20000$ & $10000$ & 7.800(9)  & 139.4(7)   &  146.7(7)&164.1(1.7)\\
      & 64 & $20000$ &  $5000$ & 7.917(4)  & 139.7(8)   &  148.2(7)  & 158(2) \\
      &128 & $10000$ &  $5000$ & 7.924(2)  & 138.8(1.0) & 147.1(1.1) & 157(3) \\
0.282 & 64 & $15000$ & $20000$ & 9.331(6)  & 207.0(1.0) & 220(2)     & 238(3) \\
      &128 & $ 5000$ & $20000$ & 9.346(5)  & 205.5(1.6) & 218.8(1.3) & 232(4) \\
0.283 & 64 & $20000$ & $20000$ &11.551(10) & 342.8(2.1) & 361.0(1.6) & 402(4) \\
0.284 &128 & $ 5000$ & $50000$ &15.837(16) & 716(6)     & 753(9)     & 842(19)\\
\end{tabular}
\end{ruledtabular}
\end{table}

Some results are reported in Table~\ref{tab:tabHT}. 
Since we are interested in infinite-volume quantities, we must be sure
that finite-size effects are negligible. A detailed check is performed 
at $\beta = 0.281$, where we can compare simulation results at
different values of $L$, corresponding to 
$\xi(\beta,L)/L \approx 0.24$, 0.12, and 0.06. 
No scaling corrections are observed in $\tau_{\rm int}(0)$ within the 
quoted errors, and thus, for each $\beta$, we assume that the
estimate of $\tau_{\rm int}(0)$ 
for the largest lattices is an infinite-volume result. 
Also $\tau_{{\rm exp},1}(0)$ apparently does not show finite-size effects. 
On the other hand,  $\tau_{{\rm exp},2}(0)$ is clearly decreasing as 
$L$ increases. 
This indicates that finite-size effects on $\widetilde{G}(k,t)$ increase
with $t$, a result that we will check explicitly below, considering the 
correlation function.

\subsection{Dynamic structure factor} \label{sec3.4}

We first use the estimates of the autocorrelation times to obtain
an estimate of $z$. Since the model is approximately improved,\cite{HPPV-07}
the scaling corrections proportional to $(\beta_c - \beta)^{\omega\nu}$,
$\omega = 0.29(3)$ are suppressed. Thus, the leading scaling corrections
behave as $(\beta_c - \beta)^{\omega_2\nu}$, where $\omega_2=0.82(8)$
is the next-to-leading correction-to-scaling exponent. Hence,
$\tau(\beta)$ behaves as 
\begin{equation}
\tau(\beta) \approx 
    c (\beta_c - \beta)^{-z \nu} (1 + b (\beta_c - \beta)^{\Delta_2}
   +\cdots),
\end{equation}
where $\Delta_2 = \nu \omega_2 = 0.56(6)$.
Thus, we fit the data to 
\begin{equation}
  \ln \tau(\beta) = - z \nu \ln (\beta_c - \beta) + a + 
    b (\beta_c - \beta)^{\Delta_2},
\end{equation}
setting $\beta_c = 0.2857431(3)$. 
\cite{HPPV-07,HPV-07}
If we fit $\tau_{\rm int}(0)$, including only the data satisfying 
$\beta \ge \beta_{\rm min}$, we obtain
$z \nu = 1.64(3), 1.59(4), 1.62(8)$ for $\beta_{\rm min} = 0.275$,
0.278, 0.280. The results are stable with $\beta_{\rm min}$ 
and allows us to estimate $z \nu = 1.61(6)$ that includes 
all estimates with their error bars.
If we now use\cite{HPPV-07}
$\nu = 0.683(2)$, we obtain 
\begin{equation}
   z = 2.36(9),
\end{equation}
which is in perfect agreement with the estimate $z = 2.35(2)$ obtained at the 
critical point.\cite{HPV-07} As a check we have repeated the 
analysis by using $\tau_{{\rm exp},1}(0)$. We obtain 
$z \nu = 1.59(3), 1.56(5), 1.47(10)$ for $\beta_{\rm min} = 0.275$,
0.278, 0.280, which are essentially consistent with the estimates
obtained above.

\begin{figure*}[tb]
\centerline{\psfig{width=9truecm,angle=-90,file=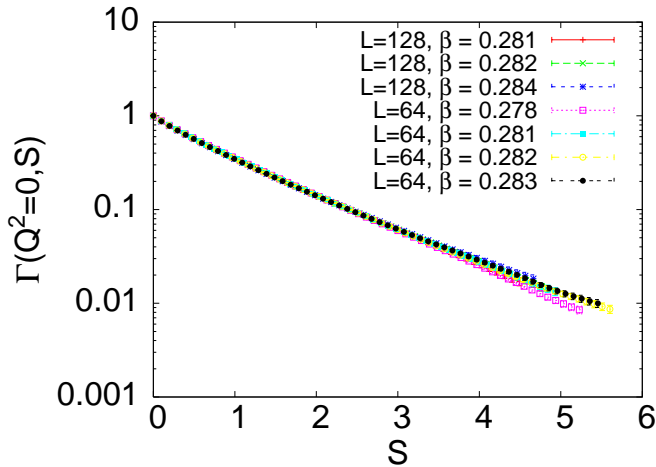}}
\vspace{2mm}
\caption{(Color online)
Scaling function 
$\widetilde{G}(0,t)/\widetilde{G}(0,0)$ as a function of $S$, for 
different values of $\beta$ and $L$.
}
\label{fig:Gkt-MC-keq0}
\end{figure*}

\begin{figure*}[tb]
\centerline{\psfig{width=9truecm,angle=-90,file=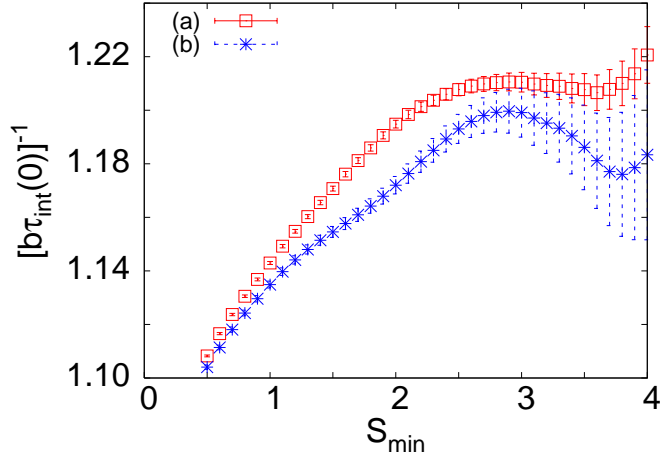}}
\vspace{2mm}
\caption{(Color online)
Estimates of $[b\tau_{\rm int}(0)]^{-1}$, which converges to 
$f_{\rm exp}(0)$ in the critical limit. The coefficient $b$ is 
obtained by fitting
$\widetilde{G}(0,t)/\widetilde{G}(0,0)$, as described in the text,
see Eq.~(\ref{fit-Gtilde}). In fit (a) 
we consider together the data corresponding to
($L=64$, $\beta = 0.278$), ($L=128$, $\beta = 0.281$), and 
($L=128$, $\beta = 0.282$). In fit (b) we only consider the 
results obtained for $L=128$, $\beta = 0.281$.
}
\label{fig:tauexp-tauint-ratio}
\end{figure*}

Let us now consider $\widetilde{G}(k,t)$. Let us first focus on the 
case $k = Q = 0$. Numerical results are reported in 
Fig.~\ref{fig:Gkt-MC-keq0} vs $S\equiv t/\tau_{\rm int}(0)$. Scaling and 
finite-size corrections are small and indeed all data fall approximately onto 
a single curve. Some deviations are only observed for $S\gtrsim 3$,
indicating that finite-size corrections
increase with $S$. Let us now consider the large-$S$ behavior
and let us estimate the universal ratio $f_{\rm exp}(0)$. The data 
show a reasonably good exponential behavior so that we can assume that 
we are considering values of $t$ that are much before the region 
in which $\widetilde{G}(0,t)$ shows the Griffiths tail.
We perform fits of the form
\begin{equation}
\ln {\widetilde{G}(0,t)\over \widetilde{G}(0,0)} = a - b t,
\label{fit-Gtilde}
\end{equation}
including each time only data in the range
$S_{\rm min}\le S \le S_{\rm max}\approx 5$. The fit parameter
$b$ provides an estimate of $f_{\rm exp}(0)$: in the critical 
limit $f_{\rm exp}(0) = [b\tau_{\rm int}(0)]^{-1}$. Since 
finite-size corrections are important, 
we only consider data with small $\xi/L$. 
In Fig.~\ref{fig:tauexp-tauint-ratio} we report the results 
corresponding to two sets of data. In fit (a) we consider three 
sets of results: those corresponding to $L=64$, $\beta = 0.278$ and 
those with $L=128$ and $\beta = 0.281,0.282$. Correspondingly,
we have $\xi(\beta,L)/L = 0.088,0.061,0.073$, respectively. 
In fit (b) we only use the lattice with the smallest value of 
$\xi/L$ available: $L=128$ and $\beta = 0.281$. The results of fit
(a) become independent of $S_{\rm min}$ for $S_{\rm min}\gtrsim 2.5$
and give $f_{\rm exp}(0) = 1.21(1)$.
Fit (b) is less stable and a plateau is less evident. They hint at a 
lower value for the ratio, varying between 1.20 (at $S_{\rm min} = 3$)
and $1.18$ (at $S_{\rm min} = 4$), though with a large statistical error.
We have also analyzed the data corresponding to lattices with 
larger $\xi/L$, finding larger values of $[b\tau_{\rm int}(0)]^{-1}$.
This indicates that this quantity decreases with decreasing $\xi(\beta,L)/L$
and thus the difference obtained between fits (a) and (b) may be a real
finite-size effect. For this reason our final result corresponds to 
fit (b). We quote
\begin{equation}
f_{\rm exp}(0) = 1.19(3),
\label{tauratio_Q20}
\end{equation}
where the error has been chosen conservatively, in order to include the 
result of fit (a) with its error. This result is very close to 
the one-loop FT estimate. Eq.~(\ref{tauexp-tauint}) gives 
$f_{\rm exp} \approx 1.168$ for $\epsilon=1$.

Finally, we provide an interpolation of our numerical data. The curves reported
in Fig.~\ref{fig:Gkt-MC-keq0} are well fitted by a function of the form
\begin{equation}
\Gamma(0,S) = e^{-\kappa S} + 
\sum_{k=1}^4 a_k {(c S)^k\over 1 + (c S)^4 e^{\kappa S}}.
\label{G0S-param}
\end{equation}
The constant $\kappa$ has been fixed by using Eq.~(\ref{tauratio_Q20}): 
we take $\kappa = 0.840 \approx 1/1.19$.
All other constants have been obtained 
by a fit of the data for $0\le S \le 5$. We obtain 
$c = 1.69$, $a_1 = -0.23825$, $a_2 = 0.16430$, $a_3 = 0.13261$, 
$a_4 = -0.28028$. As a check of this parametrization we verify the 
normalization conditions (\ref{norm-Gamma0}). The first condition 
is satisfied exactly, the second one to very good precision: the integral
between 0 and infinity 
of $\Gamma(0,S)$ as given by the parametrization (\ref{G0S-param})
is equal to 1.0033.

\begin{figure*}[tb]
\centerline{\psfig{width=9truecm,angle=-90,file=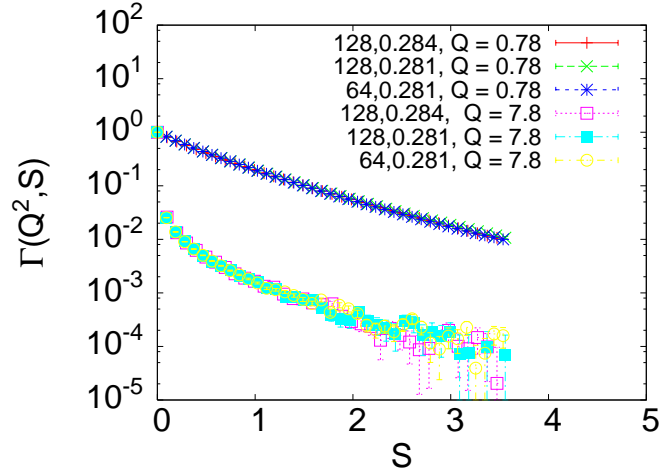}}
\vspace{2mm}
\caption{(Color online)
Scaling function $\Gamma(Q^2,S)$ 
as computed numerically for two values of  
$Q^2$. For each $Q$, we report results corresponding to three different 
values of $\beta$ and $L$.
}
\label{fig:Gkt-MC-knot0}
\end{figure*}

Let us now consider $\widetilde{G}(k,t)$ for $k\not=0$. Again, let us 
first discuss the finite-size and scaling corrections. For this purpose,
we must compare $\widetilde{G}(k,t)$ for different values of $\beta$
and $L$, but at the same value of $Q \equiv k \xi$. Since the momenta
accessible on a finite lattice of size $L$ are quantized and therefore
estimates are obtained only for $Q = 2 \pi n \xi/L$, $n$ integer, for each $t$
we should interpolate the numerical data as we did in Sec.~\ref{sec2.3}.
However, by a fortunate accident, such an interpolation is not needed here.
Indeed, the lattice with $L = 64$, $\beta = 0.281$  and that 
$L = 128$, $\beta = 0.284$ have both $\xi(\beta,L)/L = 0.1237(1)$. 
Moreover, for $L = 128$, $\beta = 0.281$, $\xi/L$ is exactly 1/2 (within the
small statistical errors) of the 
previous value. Thus, results with the same $k$ for the first two systems and 
those with $2k$ for the third one 
correspond quite precisely to the same value of 
$Q$. In Fig.~\ref{fig:Gkt-MC-knot0} we report results corresponding to 
$Q = 2\pi\times 0.1237 \approx 0.78$ and 
$Q = 20\pi\times 0.1237 \approx 7.8$. All results fall again onto a single
curve for both values of $Q$. Finite-size and scaling corrections 
are apparently negligible in this range of values of $Q$ and $S$. 
This result should be compared with what we observed for the static structure 
factor in Sec.~\ref{sec2.3}. There, a good scaling behavior was only observed 
at fixed $\widehat{Q}$ and not at fixed $Q$. Here instead, scaling corrections
at fixed $Q$ are quite small; the behavior at fixed $\widehat{Q}$ is 
actually slightly worse. 

\begin{figure*}[tb]
\centerline{\psfig{width=9truecm,angle=-90,file=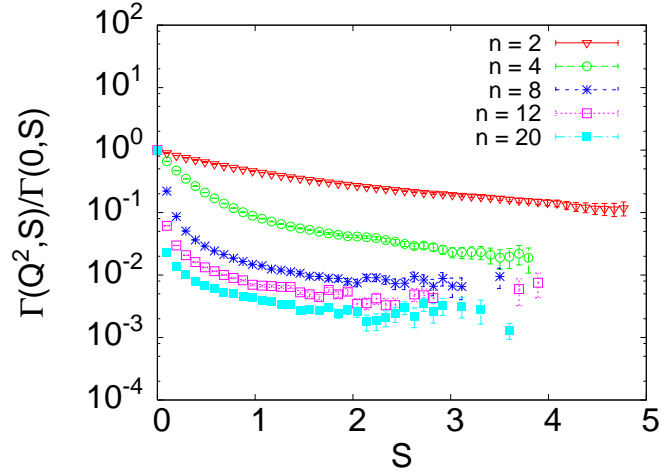}}
\vspace{2mm}
\caption{(Color online)
Ratio $\widetilde{G}(k,t)/\widetilde{G}(0,t) \approx 
\Gamma(Q^2,S)/\Gamma(0,S)$ 
obtained at $\beta = 0.282$, $L=128$, versus $S$. 
Results correspond to different
values of $Q^2 = 4\pi^2 n^2\xi^2/L^2$. The values 
$n = 2,4,8,12,20$ correspond to
$Q^2 \approx 0.84,3.37,13.5, 30.3, 84$.
}
\label{fig:GQoverG0-1}
\end{figure*}

An important prediction of the FT analysis is that $\Gamma(Q^2,S)$ decays 
with the same rate for all values of $Q$. To check this prediction we consider
the ratio $\widetilde{G}(k,t)/\widetilde{G}(0,t)$, which converges to 
$\Gamma(Q^2,S)/\Gamma(0,S)$ in the scaling limit. For $S\to \infty$, this 
quantity should behave as 
\begin{equation}
S^\alpha \exp\left[ -S (f_{\rm exp}(Q^2)^{-1} - f_{\rm exp}(0)^{-1})\right],
\end{equation}
where $\alpha$ is some critical exponent. 
Field theory
predicts $f_{\rm exp}(Q^2) = f_{\rm exp}$ independent of $Q$, so that 
we expect $\Gamma(Q^2,S)/\Gamma(0,S)$ to behave as $S^\alpha$ for large
$S$, without exponential factors. Thus, if field theory is correct, 
$\ln [\widetilde{G}(k,t)/\widetilde{G}(0,t)]$ should become constant
as $t$ increases, apart from possible slowly varying logarithmic corrections.
In Fig.~\ref{fig:GQoverG0-1} we show this ratio
for the lattice with $\beta = 0.282$, $L=128$, which has been chosen because
of its relatively small errors up to $S\approx 4$. The results for 
$\beta = 0.284$, $L=128$, which are more asymptotic and give access to larger
values of $Q$, are more noisy. The plot shows that the MC data 
are consistent with the FT prediction. Note that the constant behavior 
is observed better for larger values of $Q$. This is in agreement with 
the FT results shown in Fig.~\ref{fig:Gkt-pert} and can be understood 
qualitatively quite easily. Roughly, at one loop $\Gamma(Q^2,S)$ is the 
sum of two terms, 
\begin{equation}
a e^{- S} + b e^{-(1 + Q^2) S}
\end{equation}
(we neglect here 
additional powers of $Q$ and $S$), so that the ratio we 
are considering corresponds to $(a + b e^{- Q^2 S})/(a + b)$. 
Thus, the ratio approaches a constant with corrections of order $e^{- Q^2 S}$.
For large $Q^2$ 
they die out fast, and thus a constant behavior is observed for 
small values of $S$. 

\begin{figure*}[tb]
\begin{center}
\begin{tabular}{cc}
\psfig{width=8truecm,angle=-90,file=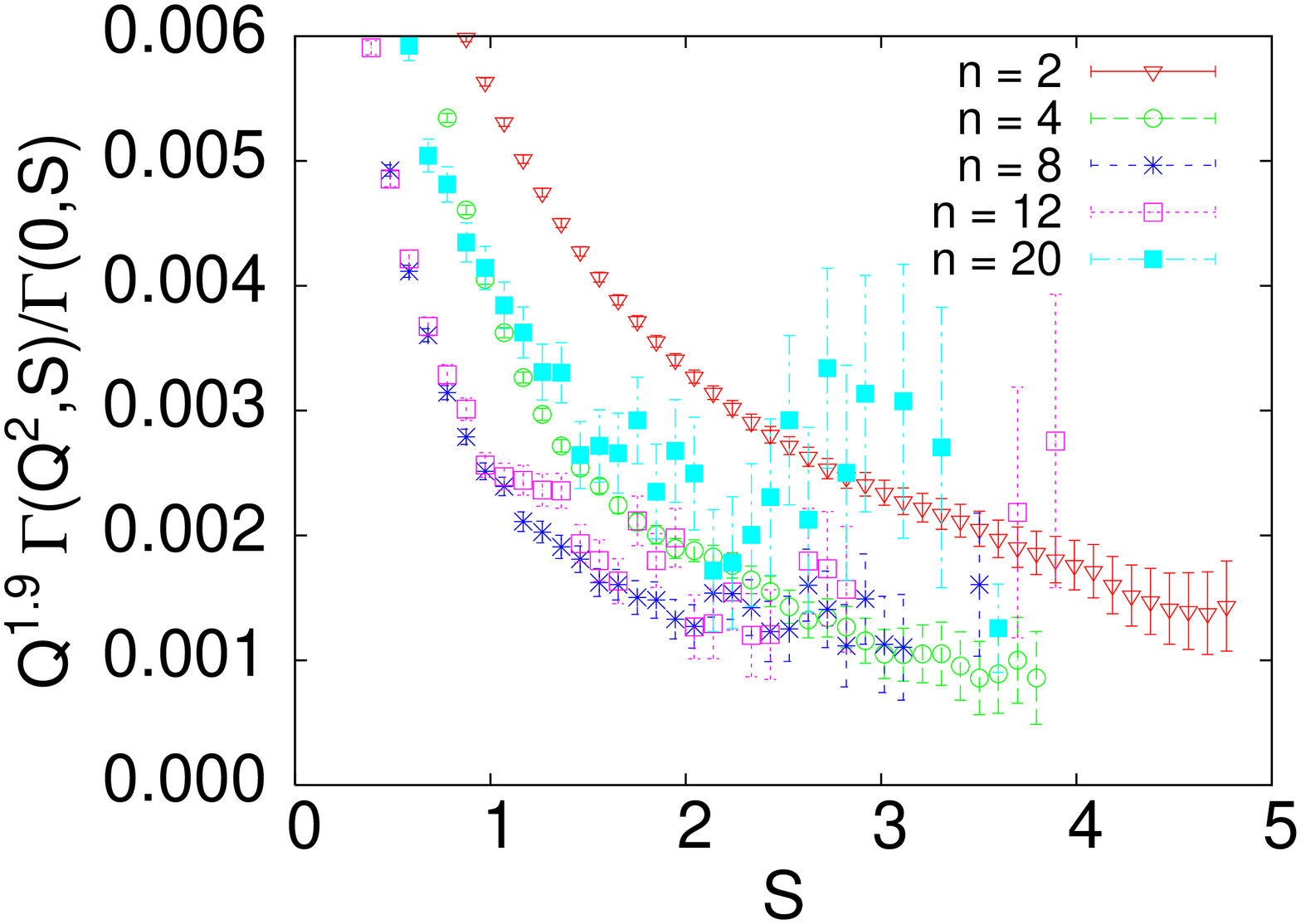} &
\psfig{width=8truecm,angle=-90,file=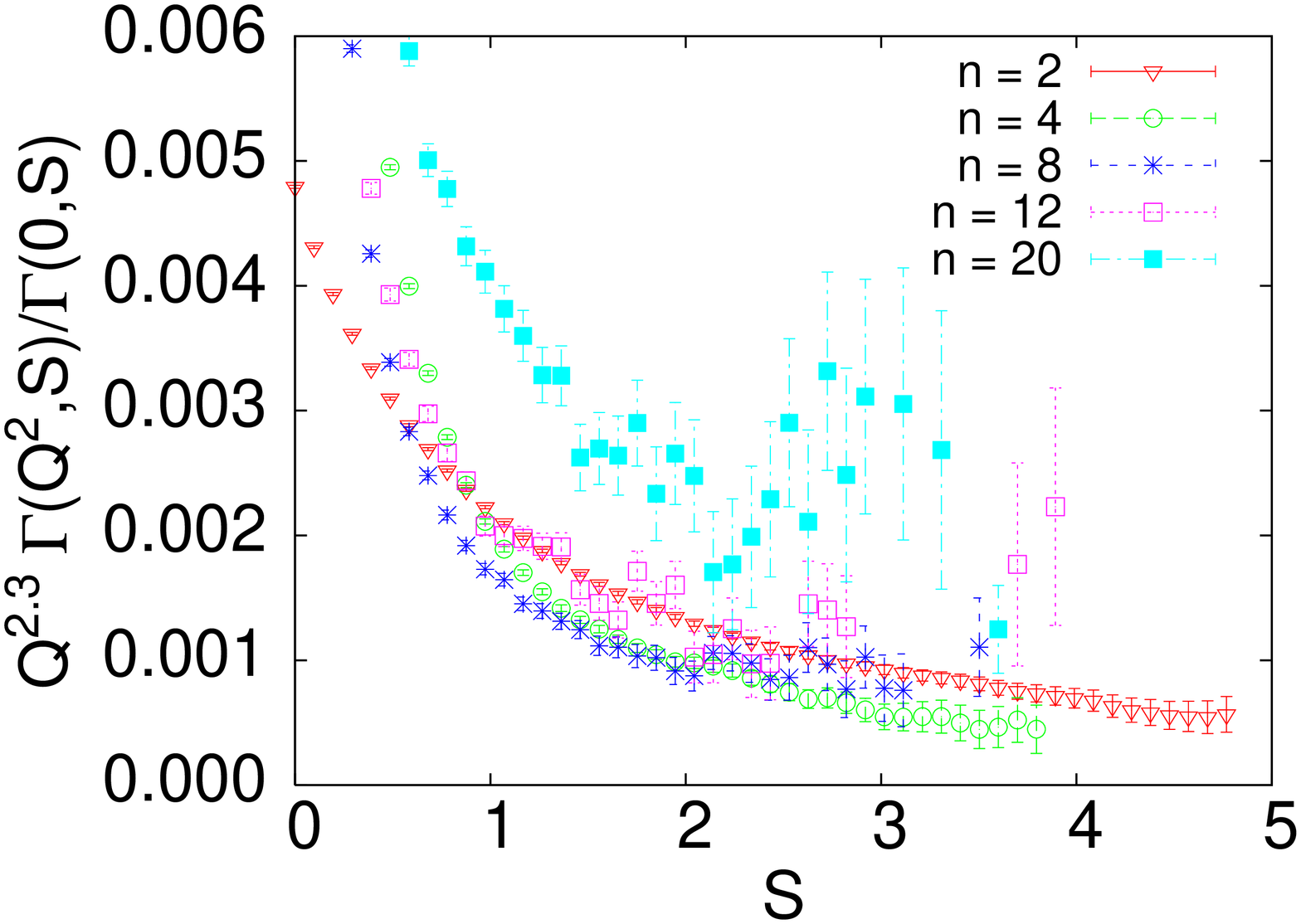} 
\end{tabular}
\end{center}
\vspace{2mm}
\caption{(Color online)
Ratio 
$Q^{\zeta}\widetilde{G}(k,t)/\widetilde{G}(0,t) \approx 
 Q^{\zeta} \Gamma(Q^2,S)/\Gamma(0,S)$ 
obtained at 
$\beta = 0.282$, $L=128$, versus $S$. Results correspond to different
values of $Q^2 = 4\pi^2 n^2\xi^2/L^2$. The values 
$n = 2,4,8,12,20$ correspond to
$Q^2 \approx 0.84,3.37,13.5, 30.3, 84$. The figure on the left corresponds to 
$\zeta = 1.9$, that on the right to $\zeta = 2.3$. 
}
\label{fig:GQoverG0-2}
\end{figure*}

While the decay rate of $\Gamma(Q^2,S)$ is independent of $Q^2$, the amplitude
decreases rapidly with $Q^2$. For large $Q^2$ we expect
the behavior $\Gamma(Q^2,S) \sim S^a Q^{-\zeta} \exp(- \kappa S)$,
where $\kappa = 1/f_{\rm exp}$, see Eq.~(\ref{Gamma-largeS}). 
We wish now to obtain a rough estimate of 
the exponent $\zeta$. For this purpose we 
take the data that appear in Fig.~\ref{fig:GQoverG0-1} and we multiply
them by $Q^{\zeta}$, trying to fix $\zeta$ 
in such a way to obtain a good collapse of 
the data. In Fig.~\ref{fig:GQoverG0-1} we report the scaled results 
corresponding to two different values of $\zeta$. If we try to have a good 
collapse of the data corresponding to $n=4,8,12$ the best result is 
obtained for $\zeta = 2.3$. However, the data with $n = 20$ behave in
a significantly different way. If we try to include also the data with 
$n = 20$, the quality of the collapse worsens and the best result is 
obtained for $\zeta = 1.9$. These results indicate that $\zeta\approx 2$ 
(but with a large error),
so that $\Gamma(Q^2,S)$ behaves roughly as $Q^{-2} S^a \exp(- \kappa S)$. 
It is interesting
to observe that this is exactly the behavior predicted 
close to four dimensions by perturbation theory.

\begin{table}
\squeezetable
\caption{
Numerical values of the coefficients appearing in 
the interpolation formula (\ref{interp-gQ2}).
}
\label{tabcoeff}
\begin{ruledtabular}
\begin{tabular}{rrrrr}
&$Q^2 = 0.842$& $Q^2 = 5.262$& $Q^2 = 21.05$ & $Q^2 = 47.36$ \\ \hline
$a_0$ &  $-$2.30388 &  0.0972492 & $-$0.1390160  & $-$0.1482217   \\
$a_1$ &   1.85440 & $-$0.1867741 & $-$0.0166780  &  0.0609449  \\
$a_2$ &  $-$1.03426 &  0.2245075 &  0.0882145  & $-$0.0108184  \\
$a_3$ &  $-$1.04434 & $-$0.0670895 & $-$0.0483811  & $-$0.0040681  \\
$a_4$ &   2.29488 & $-$0.0788428 &  0.1494524  &  0.1510844  \\
$d_0$ &   3.30388 &  0.9027518 &  1.1390167  &  1.1482211  \\
$d_1$ &  $-$1.87228 & $-$0.5008092 & $-$13.808512  & $-$19.439356  \\
$d_2$ &   2.03391 &  12.846119 &  114.32824  &  182.53584  \\
$d_3$ &  $-$0.64676 & $-$27.878988 & $-$319.65479  & $-$587.00842  \\
$d_4$ &   0.13949 &  58.784027 &  519.97023  &  1021.6385  \\
$\kappa_2$ &   1.95    &   9.08  &   10.46     &    12.40       \\
$c$   &   2.000   &  1.124  &   1.087     &    1.282      \\
\end{tabular}
\end{ruledtabular}
\end{table}

Finally, we determine an interpolation formula for $\Gamma(Q^2,S)$. 
We find that all data are well fitted by taking 
\begin{equation}
\Gamma(Q^2,S) = a_0 e^{-\kappa_1 S} + 
    \sum_{k=1}^4 a_k {(c S)^k\over 1 + (c S)^4 e^{\kappa_1 S}}
     + \sum_{k=0}^4 d_k S^k e^{-\kappa_2 S},
\label{interp-gQ2}
\end{equation}
fixing $\kappa_1 = 0.840$.  The results of the fits for a few chosen values of 
$Q^2$ are reported in Table \ref{tabcoeff}. 
We have not required $a_0 + d_0 = 1$,
a condition that follows from $\Gamma(Q^2,S=0) = 1$, but we have verified that 
the results satisfy this condition quite precisely.
By using a linear interpolation, the results we report should allow the reader 
to determine $\Gamma(Q^2,S)$ for any $Q$ in the range $0\le Q^2 \le 50$ 
with reasonable precision. We stress that this interpolation formula 
only represents a compact expression that summarizes the numerical results.
The chosen parametrization has indeed a purely phenomenological value.

\begin{figure*}[tb]
\centerline{\psfig{width=9truecm,angle=-90,file=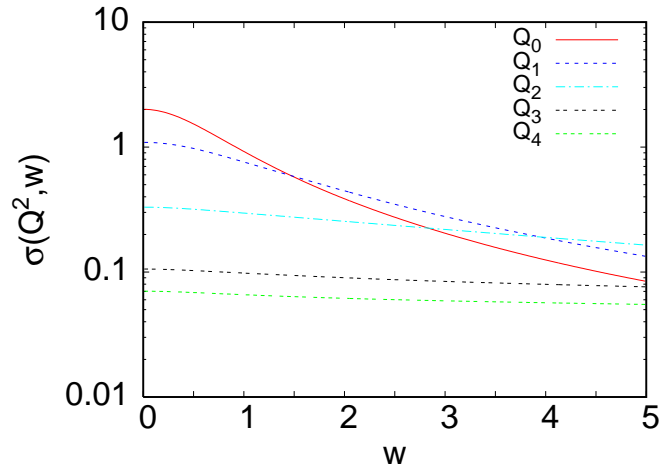}}
\vspace{2mm}
\caption{(Color online)
Numerical estimate of the scaling function $\sigma(Q^2,w)$ as a function of 
$w \equiv \omega\tau_{\rm int}(0)$, as obtained
by integrating the interpolating expression
(\ref{interp-gQ2}). The values of the momenta are:
$Q_0 = 0$, $Q_1^2 = 0.842$, $Q_2^2 = 5.262$, $Q_3^2 = 21.05$,
$Q_4^2 = 47.36$.
}
\label{fig:Gom-num}
\end{figure*}

From these expressions it is easy to determine the scaling function 
$\sigma(Q^2,w)$ related to the dynamic structure factor.
In Fig.~\ref{fig:Gom-num} we plot the scaling function $\sigma(Q^2,w)$ 
as obtained by integrating the interpolating function determined above. 
We report it for the same values of $Q^2$ that appear in Table~\ref{tabcoeff}.
The qualitative behavior is in full agreement with the FT prediction,
compare with Fig.~\ref{fig:Gom-pert}. 
Quantitatively, perturbation theory is also
reasonably predictive. For $Q = 0$ and $w < 5$, the relative differences 
between the FT and the numerical expression are less than 2\%. For larger
values of $Q^2$ differences increase: for instance, for the values of 
$Q$ that appear in Fig.~\ref{fig:Gom-num}, field theory predicts
$\sigma(Q^2,0) \approx 2$, 1.09, 0.32, 0.091, 0.041, while we obtain numerically
$\sigma(Q^2,0) \approx  2$, 1.09, 0.33, 0.105, 0.072. These discrepancies are 
probably the fault of both field theory---after all, we are at one loop---and 
of the numerical results---for large $Q$ the data are noisy and the 
estimates have a large error. In any case this comparison 
indicates that, up to $Q = 5$,  errors are under control and the reported 
expressions are precise enough for all practical purposes.



\appendix

\section{Perturbative results for the static structure factor} \label{AppA}

The static behavior of Ising systems with random dilution can 
be studied starting from the 
Landau-Ginzburg-Wilson Hamiltonian~\cite{GL-76}
\begin{equation}
{\cal H} = \int d^d x 
\left[ {1\over 2}(\partial_\mu \varphi)^2 + {1\over 2} r \varphi^2 
+ {1\over 2} \psi \varphi^2  + {1\over 4!} g_0 \varphi^4 \right],
\label{Hphi4ran}
\end{equation}
where $\psi(x)$ is a spatially uncorrelated random field with Gaussian
distribution
\begin{equation}
P(\psi) = {1\over \sqrt{4\pi w}} \exp\left[ - {\psi^2\over 4 w}\right].
\end{equation}
Using the standard replica trick, it is possible to replace the quenched
average with an annealed one.  As a result of this procedure, one can
investigate the static critical behavior of RDI systems by applying standard
FT methods to the Hamiltonian~\cite{GL-76}
\begin{equation}
{\cal H}_{\rm replica} = \int d^d x 
\Bigl\{ \sum_{i}{1\over 2} \left[ (\partial_\mu \phi_{i})^2 + 
         r \phi_{i}^2 \right]  
 + \sum_{ij} {1\over 4!}\left( u_0 + g_0 \delta_{ij} \right)
          \phi^2_{i} \phi^2_{j} 
\Bigr\},
\label{Hphi4}
\end{equation}
where $i,j=1,...N$ and $u_0=  -6w$. RDI results are obtained by taking the 
limit $N\to 0$.

\subsection{$\sqrt{\epsilon}$-expansion results}
\label{appsqe}

The scaling function $g(Q^2)$ can be determined by using the results 
reported in Ref.~\onlinecite{CPRV-98}. We obtain 
the expansion 
\begin{eqnarray}
g(y)^{-1}&=&
1+y-\epsilon\frac{1}{53} \psi_2 (y)+
18\frac{\sqrt{6/53}}{2809}[24+7\zeta(3)] \epsilon^{3/2} \psi_2(y)
+O(\epsilon^2)
\nonumber \\
&=& 1+y-0.0188679 \epsilon \psi_2(y)+ 0.069887\epsilon^{3/2} \psi_2(y)
+O(\epsilon^2)\,,
\label{2pteps}
\end{eqnarray}
where $\psi_2 (y)$ is the two-loop contribution defined in 
Ref.~\onlinecite{CPRV-98}.
Note that the only relevant three-loop diagram contributes only at order 
$\epsilon^{5/2}$ (hence, at five loops), since it is proportional to 
\begin{equation}
{1\over27} (u + 3 v) (4 u + 3 v)^2,
\end{equation}
and, at the fixed point, $4 u^* + 3 v^*$ is of order $\epsilon$ and not of 
order $\sqrt{\epsilon}$.

The expansion of $\psi_2(y)$ for small $y$ can be found in 
Ref.~\onlinecite{CPRV-98}. 
It allows us to obtain the expansions
\begin{eqnarray}
c_2&=&0.000141891 \epsilon  - 0.000525567  \epsilon^{3/2} + O(\epsilon^2), 
\nonumber \\
c_3&=&-3.62134 \cdot 10^{-6} \epsilon  + 0.0000134135  \epsilon^{3/2}
+ O(\epsilon^2), \nonumber \\
c_4&=& 1.53623 \cdot 10^{-7} \epsilon  - 5.69021 \cdot 10^{-7} \epsilon^{3/2}
+ O(\epsilon^2), \nonumber \\
c_5&=&-8.28575 \cdot 10 ^{-9}\epsilon  + 3.06905 \cdot 10^{-8}  \epsilon^{3/2}
+ O(\epsilon^2). 
\label{exp-cn-sqrte}
\end{eqnarray}
The expansion of $\psi_2(y)$ for large values of $y$ is \cite{Bray-76}
\begin{equation}
\psi_2(y)=-\frac{1}{4} y\log y+2 Q_0 y-\frac{3}{4} \log^2y+2 Q_1+\dots
\end{equation}
with $Q_0\approx 0.507826$ and $Q_1 \approx 0.1289$.
Matching the large-momentum expansion of Eq.~(\ref{2pteps}) with the 
Fisher-Langer behavior (\ref{eq:FL})
we obtain the expansions of the coefficients $C_i$: 
\begin{eqnarray}
&&C_1 = 1 + 0.0191632 \epsilon - 0.0709809 \epsilon^{3/2} + 
O(\epsilon^2), 
\nonumber  \\ 
&&C_2 = -1/2 - 0.757042 \epsilon^{1/2} 
+ 1.34297 \epsilon + c_{2,3} \epsilon^{3/2} + 
O(\epsilon^2), 
\nonumber \\ 
&&C_3 = -1/2 + 0.757042 \epsilon^{1/2} -
1.35726 \epsilon + c_{3,3} \epsilon^{3/2} + 
O(\epsilon^2), \label{lqc}
\end{eqnarray}
where $c_{2,3}+c_{3,3}=0.052964$.
Setting $\epsilon = 1$, we obtain $C_1 \approx 0.95$, $C_2 + C_3 = -0.96$.

\subsection{Massive zero-momentum results} \label{AppA.2}

We have determined the low-momentum behavior of $g(Q^2)^{-1}$ in the 
MZM scheme by using the perturbative results of Ref.~\onlinecite{CPRV-98}.
The four-loop expansions of the first few coefficients $c_n$ are:
\begin{eqnarray}
c_2=&&  -{1\over 6480} u^2 - {1\over 2430} u v - {2\over 10935} v^2 
- 0.0000120404 u^3  - 0.0000481617 u^2 v 
\nonumber \\
&& - 0.0000481617 u v^2 
- 0.0000142701 v^3 - 0.00000718972 u^4   - 0.0000383452 u^3 v 
\nonumber\\ 
&&  - 0.0000617074 u^2  v^2  
- 0.0000379116 u v^3  - 0.00000842479   v^4,
\nonumber  \\
c_3=&&  {1\over 122472} u^2 + {1\over 45927} u v + {4\over 413343} v^2 
- 0.00000281924 u^3  - 0.0000112769 u^2 v 
\nonumber \\
&& - 0.0000112769 u v^2 
- 0.00000334132 v^3 + 0.00000660026 u^4   + 0.00000352014 u^3 v 
\nonumber\\ 
&&  + 0.00000562891 u^2  v^2  
+ 0.00000339371 u v^3  + 0.000000754158    v^4,
\nonumber \\
c_4=&&  -{1\over 1889568} u^2 - {1\over 708588} u v - {1\over 1594323} v^2 
+ 0.000000347995 u^3  + 0.00000139198 u^2 v 
\nonumber\\
&& + 0.00000139198 u v^2 
+ 0.000000412438 v^3 - 0.000000141114 u^4  - 0.000000752609 u^3 v 
\nonumber\\ 
&&  - 0.00000119627 u^2  v^2  
- 0.000000740733 u v^3  - 0.000000164607    v^4.
\label{cnMZM}
\end{eqnarray}
The MZM renormalized quartic couplings $u$ and $v$ are normalized so that
at tree level $u=u_0/m$ and $v=v_0/m$.
Their fixed-point values
are $u^*=-18.6(3)$ and $v^*=43.3(2)$ (obtained by means of MC 
simulations~\cite{CMPV-03}), and 
$u^*=-13.5(1.8)$ and $v^*=38.0(1.5)$ (obtained by resumming the six-loop  
$\beta$-function~\cite{PV-00}). 

\section{One-loop calculation of the response and correlation functions}
\label{AppB}

The relaxational model-A dynamics is described by the stochastic Langevin
equation \cite{HH-77}
\begin{equation}
\label{lang}
\frac{\partial \varphi (r,t)}{\partial t}=-\Omega 
\frac{\delta \cal{H}(\varphi)}{\delta \varphi(r,t)}+\zeta(r,t),
\end{equation}
where $\varphi(r,t)$ is the order parameter, $\cal{H}(\varphi)$ is the
Hamiltonian (\ref{Hphi4ran}), $\Omega$ is a transport coefficient, 
and $\zeta(r,t)$ is a Gaussian random field (white noise) with correlations
\begin{equation}
\langle \zeta(r,t)\rangle = 0, \qquad
\langle \zeta(r_1,t_1) \zeta(r_2,t_2)\rangle = 
    \Omega \delta(r_1-r_2) \delta(t_1-t_2). 
\end{equation}
The correlation functions generated by the Langevin equation (\ref{lang}) at
equilibrium, averaged over the noise $\zeta$ and the quenched disorder $\psi$,
can be obtained from the FT action~\cite{DeDominicis-78}
\begin{eqnarray}
S(\varphi,\hat{\varphi}) = &&
\int d^d x \left[ 
\int d t \,\hat{\varphi} 
( \partial_t \varphi  - \Omega \Delta \varphi - 
  \Omega \hat{\varphi} + \Omega r \varphi) \right.
\label{sdyn} \\
&&\left.
+ {\Omega g_0\over 3!} \int dt \,\hat{\varphi} \varphi^3
+ {\Omega^2 u_0 \over 6} 
     \left( \int dt \,\hat{\varphi} \varphi \right)^2 \right],
\nonumber
\end{eqnarray}
where $\hat{\varphi}$ is the response field.  In this
framework, no replicas are introduced.~\cite{DeDominicis-78}
We consider the correlation function $G(x,t)$ and the response function 
$R(x,t)$ defined by
\begin{eqnarray}
&&G(x_2-x_1,t_2-t_1) = \langle \varphi(x_1,t_1) \varphi(x_2,t_2) \rangle,
\label{fttwop}\\
&&R(x_2-x_1,t_2-t_1) = \langle \hat{\varphi}(x_1,t_1) \varphi(x_2,t_2) \rangle,
\label{rft}
\end{eqnarray}
and their spatial Fourier transforms $\widetilde{G}(k,t)$ and 
$\widetilde{R}(k,t)$.  In equilibrium, 
they are not independent, but related by the 
fluctuation-dissipation theorem; for $t > 0$ they satisfy the relation 
$\partial_t G(x,t) = - \Omega R(x,t)$.
For a general introduction to the FT approach to equilibrium 
critical dynamics, see, e.g.,  Refs.~\onlinecite{FM-06,CG-05}.
Some perturbative calculations can be also 
found in Refs.~\onlinecite{pertcalc}.

\subsection{One-loop calculation} \label{AppB.1}

\begin{figure*}[tb]
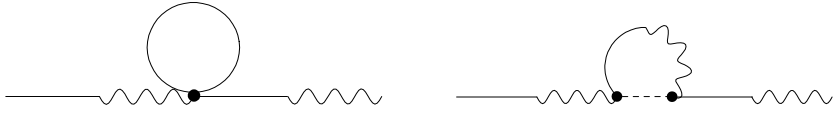

\begin{center}
\begin{tabular}{ccc}
\psfig{width=5truecm,angle=0,file=random1a.eps} & $\hphantom{????}$
\psfig{width=5truecm,angle=0,file=random1b.eps} \\
\end{tabular}
\end{center}
\vspace{-2mm}
\caption{
The one-loop graphs contributing to the response function. We indicate with 
$L_g(k,t)$ and with $L_{u}(k,t)$ the contribution of the 
graph on the left and on the right, respectively.
}
\label{graphsfig}
\end{figure*}

We first compute the response function $\widetilde{R}(k,t)$ and then use the 
fluctuation-dissipation theorem to derive the correlation function
$\widetilde{G}(k,t)$.
At one-loop we obtain in dimensional regularization 
\begin{equation}
\widetilde{R}(k,t) = \widetilde{R}_G(k,t) - {g_0\over2} L_g(k,t) - 
   {u_0\over 3} L_u(k,t) + O(u_0^2,u_0 g_0, g_0^2),
\label{oneloop}
\end{equation}
where 
\begin{equation}
\widetilde{R}_G(k,t)  =  \theta(t) \exp[-\Omega (k^2+r)t]
\label{RGauss}
\end{equation}
is the Gaussian tree-level response function, and $L_u(k,t)$ and 
$L_g(k,t)$ are the one-loop contributions, see Fig.~\ref{graphsfig}.
It is straightforward to obtain 
\begin{eqnarray}
L_g(k,t)&=& \Omega \int_{-\infty}^\infty ds \int {d^d q\over (2\pi)^d} \, 
\widetilde{R}_G(k,t-s) \widetilde{R}_G(k,s) \widetilde{G}_G(q,0)
\nonumber \\
&=& - {N_d\over \epsilon} r^{1-\epsilon/2} \Omega t \widetilde{R}_G(k,t) ,
\label{lg}
\end{eqnarray}
where $\epsilon\equiv 4-d$, $N_d \equiv 2/[\Gamma(d/2)(4\pi)^{d/2}]$, and 
$\widetilde{G}_G(k,t)$ 
is the Gaussian tree-level correlation function (\ref{Gkt-G}). 
Analogously, we obtain
\begin{eqnarray}
&&L_u(k,t)=\Omega^2 \int_{-\infty}^\infty
 ds_1 \int_{-\infty}^\infty ds_2 \int {d^d q\over (2\pi)^d} \,
\widetilde{R}_G(k,t-s_2)    \widetilde{R}_G(q,s_2-s_1) \widetilde{R}_G(k,s_1)
\nonumber \\
&&= {N_d\over \epsilon} (1 + \epsilon \gamma_E/2) (\Omega t)^{\epsilon/2} 
   (\Omega t k^2 - 1) \widetilde{R}_G(k,t) 
+ \theta(t)  {1\over (4\pi)^2} e^{-\Omega t r} F(\Omega t k^2) + O(\epsilon),
\label{lu}
\end{eqnarray}
where
\begin{equation}
F(x) \equiv   -1 + e^{-x}  + e^{-x} (x-1) [{\rm Ei}(x) - \gamma_E - \ln x],
\end{equation}
and ${\rm Ei}(x)$ is the exponential integral function.

Renormalizing the response function in the $\overline{\rm MS}$ scheme 
we obtain 
\begin{eqnarray}
\widetilde{R} (k,t) &=& \widetilde{R}_G(k,t)  -
    {g\over 32 \pi^2 } \Omega t r\ln r \widetilde{R}_G(k,t)
\\
   && -{u\over 48\pi^2} (\gamma_E + \ln \Omega t)
         (\Omega t k^2 - 1) \widetilde{R}_G(k,t)  -
         {u\over 48\pi^2} \theta(t) e^{-\Omega t r} F(\Omega t k^2),
\label{Rren-eq}
\end{eqnarray}
where $r,\Omega,u,g$ are renormalized parameters (note that we use 
the same symbols $r$ and $\Omega$ for both the bare and the renormalized
parameters, since no confusion can arise). The final expression is obtained by 
setting $g$ and $u$ equal to their fixed-point values:
\begin{equation}
g^* = 32 \pi^2 \sqrt{ {6\epsilon\over 53}}, \qquad 
u^* = - 24 \pi^2 \sqrt{ {6\epsilon\over 53}}.
\label{FP-RIM}
\end{equation}
The correlation 
function $\widetilde{G}(k,t)$ for $t > 0$ can be obtained from 
\begin{equation}
\widetilde{G}(k,t) = \Omega \int_t^\infty \widetilde{R} (k,t),
\end{equation}
which follows from the fluctuation-dissipation theorem and the fact
that $\widetilde{G}(k,t)\to 0$ as $t\to \infty$.
A straightforward calculation gives (in the following we always assume $t > 0$)
\begin{eqnarray}
{\widetilde{G}(k,t)\over \widetilde{G}(k,0)} &=&
  e^{-\Omega t (k^2 + r)} - {g\over 32 \pi^2} \Omega t r (\ln r)
   e^{-\Omega t (k^2 + r)}
\nonumber \\
  && - {u\over 48 \pi^2} (\gamma_E + \ln \Omega t) \Omega t k^2
     e^{-\Omega t (k^2 + r)}
    - {u\over 48 \pi^2} \left ( e^{-\Omega t (k^2 + r)} - e^{-\Omega t r}
    \right)
\nonumber \\
 && - {u\over 48 \pi^2} \left(\Omega t k^2 - {r\over k^2 + r}\right)
   e^{-\Omega t (k^2 + r)}
    \left({\rm Ei}(\Omega t k^2) - \gamma_E - \ln  \Omega tk^2\right)
\nonumber \\
&&  - {u\over 48 \pi^2} {r\over k^2 + r}
   \left[{\rm Ei}(-\Omega t r) - e^{-\Omega t (k^2 + r)} (\gamma_E +
    \ln \Omega t r )\right] ,
\label{Gratio-1}
\end{eqnarray}
where 
\begin{equation}
\widetilde{G}(k,0) = {1\over k^2 + r} -
    {1\over 32 \pi^2} (g + 2 u/3) {r \ln r\over (k^2 + r)^2}.
\end{equation}
It is easy to derive the critical correlation function
at one-loop order from Eq.~(\ref{Gratio-1}). 
Taking the limit $r \to  0$, we obtain 
\begin{eqnarray}
{\widetilde{G}(k,t)\over \widetilde{G}(k,0)} &=&
  e^{-\Omega t k^2} 
  - {u\over 48 \pi^2} (\gamma_E + \ln \Omega t) \Omega t k^2
     e^{-\Omega t k^2}
    - {u\over 48 \pi^2} \left ( e^{-\Omega t k^2} - 1 \right)
\nonumber \\
 && - {u\over 48 \pi^2} \Omega t k^2
   e^{-\Omega t k^2}
    \left({\rm Ei}(\Omega t k^2) - \gamma_E - \ln  \Omega tk^2\right).
\end{eqnarray} 
In the critical limit the correlation function should only depend on the 
scaling variable $K^2 \equiv k^2 (\Omega t)^{2/z}$. Keeping into account that 
$z = 2 - u/(2 4\pi^2)$ at one loop, we obtain at this order
(we set $u = u^*$)
\begin{eqnarray}
{\tilde{G}(k,t)\over \tilde{G}(k,0)} &=&
  e^{-K^2}
    - {u^*\over 48 \pi^2} \left ( e^{-K^2} - 1 \right)
    - {u^*\over 48 \pi^2} K^2 e^{-K^2}
    \left({\rm Ei}(K^2) - \ln  K^2\right).
\label{Gkt-PC}
\end{eqnarray}
This scaling function has a regular expansion in powers of 
$K^2$ for $K\to 0$, while,
for $K\to \infty$, it behaves as 
\begin{eqnarray} 
{\widetilde{G}(k,t)\over \widetilde{G}(k,0)} &\approx&
-{u^*\over 48 \pi^2 K^2} = {1\over 2 K^2} \sqrt{6\epsilon\over 53}.
\end{eqnarray}
Given Eq.~(\ref{Gkt-PC}), we can compute $G(x,t)$ at the critical 
point.  We expect the scaling behavior 
\begin{equation}
G(x,t) = (\Omega t)^{-(d + \eta - 2)/z}
F(X), \qquad\qquad 
X\equiv x (\Omega t)^{-1/z}.
\label{Gxt-scal-crit}
\end{equation}
Using Eq.~(\ref{Gkt-PC}) we obtain
\begin{equation}
F(X) = {1\over 4 \pi^2 X^2} (1 - e^{-X^2/4}) - 
       {u^*\over 48\pi^2} F_{\rm 1\, loop}(X),
\label{FX-1}
\end{equation}
with 
\begin{equation}
F_{\rm 1\, loop}(X) = {1\over 4 \pi^2 X}
\int_0^\infty dK \, J_1(KX) \left[
    e^{-K^2} - 1 + K^2 e^{-K^2}
    \left({\rm Ei}(K^2) - \ln  K^2\right) \right],
\end{equation}
where $J_1(x)$ is a Bessel function.
In the derivation we have taken into account that $u\sim\sqrt{\epsilon}$. 
It is interesting to note that $F_{\rm 1\, loop}(X)$ is not regular for 
$X\to 0$. Indeed, the explicit calculation gives
\begin{eqnarray}
F_{\rm 1\, loop}(X) &=& - {1\over 16 \pi^2} \ln {X^2\over4} + 
    O(X^2\ln X^2) .
\label{FX-2}
\end{eqnarray}
This result indicates that $F(X)$ is not regular for $X\to 0$, but behaves as 
\begin{eqnarray}
F(X) \approx a_0 + b_0 |X|^\lambda +  \cdots
\label{FX_sing}
\end{eqnarray}
where $\lambda$ is a critical exponent. Comparing this expression with 
Eqs.~(\ref{FX-1}) and (\ref{FX-2}) we obtain
\begin{eqnarray}
&& a_0 + b_0  = {1\over 16\pi^2} + O(\sqrt{\epsilon}), 
\\
&& b_0 \lambda = {u^*\over 384 \pi^4} + O(\epsilon) = 
        - {1\over 16\pi^2} \sqrt{6\epsilon\over 53} + O(\epsilon).
\end{eqnarray}
It is not possible to compute $\lambda$, $a_0$, and $b_0$ separately
at this order. A two-loop computation of the term proportional
to $\epsilon \ln^2 X$ is needed.

In the high-temperature phase, it is convenient to replace $r$ 
and $\Omega$ with the correlation length $\xi$ and the zero-momentum 
integrated autocorrelation time $\tau_{\rm int}(k=0)$,
defined in Sec.~\ref{sec2.1} and \ref{sec3.1}, respectively.
A tedious calculation gives
\begin{eqnarray}
\xi^{-2} &=& r + {1\over 32 \pi^2} (g + 2u/3) r \ln r, \\
\Omega \tau_{\rm int}(k) &=& {1\over k^2 + r} -
     {g\over 32 \pi^2} {r \ln r\over (k^2 + r)^2} +
     {u\over 48 \pi^2} {k^2 \ln r\over (k^2 + r)^2} +
     {u\over 48 \pi^2} {1\over k^2 + r}.
\label{tauintk}
\end{eqnarray}
Using these results, we obtain the one-loop expression of the 
scaling function $\Gamma(Q^2,S)$:
\begin{eqnarray}
\Gamma(Q^2,S) &=&
  e^{- S (Q^2 + 1)} - {u^*\over 48 \pi^2} S (Q^2 + 1)  e^{- S (Q^2 + 1)}
\nonumber \\
  && - {u^*\over 48 \pi^2} (\gamma_E + \ln S) S Q^2
     e^{- S (Q^2 + 1)}
    - {u^*\over 48 \pi^2} \left ( e^{-S(Q^2 + 1)} - e^{-S}
    \right)
\nonumber \\
 && - {u^*\over 48 \pi^2} \left(Q^2 S - {1\over Q^2 + 1}\right)
   e^{-S (Q^2 + 1)}
    \left({\rm Ei}(S Q^2) - \gamma_E - \ln  S Q^2 \right)
\nonumber \\
&&  - {u^*\over 48 \pi^2} {1\over Q^2 + 1}
   \left[{\rm Ei}(-S) - e^{-S (Q^2 + 1)} (\gamma_E +
    \ln S )\right],
\label{Gscal-HT}
\end{eqnarray}
where 
$u^*$ is the fixed-point value (\ref{FP-RIM}) of $u$.

It is interesting to discuss the large-$S$ and small-$S$ behavior 
of the scaling function (\ref{Gscal-HT}). For large $S$
we obtain 
\begin{equation}
\Gamma(Q^2,S) \approx 
e^{-S(1 + Q^2)} - {u^*\over 48\pi^2} {1 + Q^2\over (Q^2)^2} {e^{-S}\over S^2}
  - {u^*\over 48\pi^2} (1 + Q^2 - Q^2 \ln Q^2) S e^{-S(1 + Q^2)}.
\end{equation}
For $Q^2 = 0$ the dominant term is the last one.
In this case we can rewrite the large-$S$ behavior as 
\begin{equation}
\Gamma(Q^2,S) \approx 
\exp \left[-S \left(1 + {u^*\over 48\pi^2}\right)\right],
\end{equation}
which gives for the 
exponential autocorrelation-time scaling function [see Eq.~(\ref{fexp-Q2-def})]
\begin{equation}
f_{\rm exp}(0) = 1 - {u^*\over 48\pi^2} = 
1 + {1\over 2} \sqrt{6\epsilon\over 53} + O(\epsilon).
\label{tauexp-tauint}
\end{equation}
For $Q^2\not = 0$, the dominant term is the second one, 
so that for any
$Q^2$ the scaling function decays as $e^{-S}/S^2$. 
Thus, at this perturbative order, we obtain the result 
\begin{equation}
f_{\rm exp}(Q^2) = 1 + O(\sqrt{\epsilon}).
\label{tauexp-ratio}
\end{equation}
The correlation function $\Gamma(Q^2,S)$ decays with the same 
rate for all values of $Q$. As we discuss in Sec.~\ref{sec3.2},
this is a consequence of the loss of translational invariance in dilute 
systems.

Equation (\ref{tauexp-ratio}) should be contrasted with the 
result obtained for the integrated autocorrelation times.
Using Eq.~(\ref{tauintk}), we obtain 
\begin{equation}
{\tau_{\rm int}(k)\over \tau_{\rm int}(0)} = 
   f_{\rm int}(Q^2) = 
   {1\over Q^2 + 1},
\end{equation}
without one-loop corrections.
This shows that  $\tau_{\rm int}(k)$ decreases as $Q$ increases as it does
in the Gaussian model.

Let us now consider the limit $S\to 0$. The scaling function has an 
expansion of the form
\begin{equation}
\Gamma(Q^2,S) = 
1 + \sum_{n=1} S^n (a_n + b_n \ln S).
\end{equation}
Note the presence of terms proportional to $\ln S$. They should be 
generically expected, since in the critical limit (it corresponds 
to $Q\to\infty$ and $S\to 0$), the correlation function depends on 
\begin{equation}
k^2 (\Omega t)^{2/z} \sim Q^2 S^{2/z} \sim Q^2 S \left(1 +
    {u^*\over 48\pi^2} \ln S\right).
\end{equation}
The presence of these logarithms implies that the function $\Gamma(Q^2,S)$
is not analytic for $S\to 0$. 

It is also important to discuss the large-momentum behavior of 
$\Gamma(Q^2,S)$. For $Q^2 \to \infty$ the tree-level term 
vanishes exponentially as $e^{-S Q^2}$, while the one-loop term 
decays only algebraically, as $1/Q^2$. More precisely, for 
$Q^2 \to \infty$ we have 
\begin{equation}
\Gamma(Q^2,S) \approx 
- {u^*\over 48\pi^2} 
  \left( {e^{-S}\over S} + {\rm Ei}(-S)\right) Q^{-2}.
\label{largeQ2-HT}
\end{equation}
The presence of these slowly decaying terms implies the singularity 
of the behavior of $G(x,t)$ for $x\to 0$ and for any $t$. In the 
critical limit we expect the scaling behavior (\ref{fys}), i.e.,
$G(x,t) = \xi^{-d + 2 - \eta} F(Y^2,S)$, with $Y^2 \equiv x^2/\xi^2$.
We obtain for $Y\to 0$
\begin{equation}
F(Y^2,S) \approx {1\over 16\pi^2} 
  \left( {e^{-S}\over S} + {\rm Ei}(-S)\right) 
   \left(1 + {u^*\over 24\pi^2}\ln Y\right) + 
  u^* f_{\rm corr}(S) + \cdots
\label{FYT-pert}
\end{equation}
for $Y\to 0$, where $f_{\rm corr}(S)$ is a function of $S$. 
The presence of a term proportional to $\ln Y$ implies that 
$F(Y^2,S)$ is not analytic as $Y\to 0$, i.e. has
a behavior of the form
   $F(Y^2,S) = f_0(S) + f_\lambda(S) |Y|^\lambda + \cdots$,
where $\lambda$ is the same exponent that appears in Eq.~(\ref{FX_sing}).
Note that Eq.~(\ref{FYT-pert}) is apparently 
consistent with the assumption that 
$f_0(S) = 0$. If this were the case, we would obtain 
\begin{equation}
  \lambda = {u^*\over 24\pi^2} = - \sqrt{6\epsilon\over 53} + O(\epsilon).
\end{equation}
This result would imply $\lambda < 0$ and thus $F(Y^2,S)$ would diverge 
as $Y\to 0$ for any $S$, at least for $\epsilon$ small. 
This behavior is clearly unphysical; thus,
$f_0(S)$ should be nonvanishing.

The critical limit is obtained by taking $S\to 0$. Requiring the limiting
function to be of the form (\ref{Gxt-scal-crit}), we obtain
\begin{eqnarray}
&& f_0(S) \sim S^{(2 - \eta - d)/z},  \nonumber \\
&& f_\lambda(S) \sim S^{(-\lambda + 2 - \eta - d)/z} ,
\end{eqnarray}
for $S\to 0$. The $S$-dependent prefactor appearing in
Eq.~(\ref{FYT-pert}) behaves as $1/S$ for $S\to 0$, which is 
consistent with these expressions. 

The nonanalytic behavior of $F(Y^2,S)$ as $Y\to 0$, implies that 
$\Gamma(Q^2,S)$ should decay as a power of $Q$ as $Q\to \infty$. 
A simple calculation gives 
\begin{equation}
\Gamma(Q^2,S) \sim f_\lambda(S) Q^{2 - d - \lambda-\eta},
\label{largeQ-withc}
\end{equation}
for $Q^2 \to \infty$. The exponent $\zeta$ defined in Eq.~(\ref{f-largeQ-main})
is given by
\begin{equation}
 \zeta = \lambda + d + \eta - 2 = 2 + O(\sqrt{\epsilon}).
\end{equation}
Finally, we report $\sigma(Q^2,w)$, cf. Eq.~(\ref{defsigma}). 
A long calculation gives
\begin{eqnarray}
\sigma(Q^2,w) &=& {2\alpha\over \alpha^2 + w^2} + {u^*\over 24 \pi^2} \left\{
    {\alpha (\alpha^2 -w^2 + 2 \alpha w^2)\over
     w (\alpha^2 + w^2)^2} {\rm Arctan}\, w +
    {\alpha (\alpha^2 - w^2 - 2 \alpha) \over
   2 (\alpha^2 + w^2)^2} \ln (1 + w^2) \right.
\nonumber \\
   && \left.
    + {1\over w^2 + 1} + {(w^2 - \alpha) Q^2\over (w^2 + 1)  (\alpha^2 + w^2)}
     - {2 \alpha^3 \over (\alpha^2 + w^2)^2} \right\},
\end{eqnarray}
where $\alpha \equiv 1 + Q^2$.

For large $w$, $\sigma(Q^2,w)$ behaves as 
\begin{eqnarray}
\sigma(Q^2,w) \approx {2\alpha \over w^2} 
   \left[1 + {u^*\over 48\pi^2} (1 - \ln w)\right],
\end{eqnarray}
which is compatible with the expected behavior
$w^{-(2 - \eta + z)/z}$. 

Note also that the singularities of $\sigma(Q^2,w)$ in the complex
$w$-plane that are closest to the origin are $w=\pm i$, independently of 
$Q^2$. This is a direct
consequence of the fact we have already noticed that the large-$t$
behavior is momentum independent. As a consequence, the width of the 
structure factor does not decrease with $Q^2$ as it does in pure systems.


\begin{thebibliography}{99}

\bibitem{Belanger-00}
D. P. Belanger, Braz. J. Phys. {\bf 30}, 682 (2000);
cond-mat/0009029.

\bibitem{PV-02}
A. Pelissetto and E. Vicari,
Phys. Rept. {\bf 368}, 549 (2002).

\bibitem{FHY-03}
R. Folk, Yu. Holovatch, and T. Yavors'kii,
Uspekhi Fiz. Nauk {\bf 173}, 169 (2003)
[English translation, Phys. Usp. {\bf 46}, 175 (2003)].

\bibitem{HH-77}
P. C. Hohenberg and B. I. Halperin,
Rev. Mod. Phys. {\bf 49}, 435 (1977).

\bibitem{footBorn} 
In Born approximation the static and dynamic structure factors are
proportional to the elastic and inelastic cross section, respectively;
$k$ and $\omega$ are respectively proportional to the transferred
momentum and energy.


\bibitem{Griffiths}
R. B. Griffiths,
Phys. Rev. Lett. {\bf 23}, 17 (1969);
M. Schwartz,
Phys. Rev. B {\bf 18}, 2364 (1978).

\bibitem{v-06}
T. Vojta, J. Phys. A {\bf 39}, R143 (2006).


\bibitem{Bray-88}
A. J. Bray, 
Phys. Rev. Lett. {\bf 60}, 720 (1988).

\bibitem{DRS-88}
D. Dhar, M. Randeria, and J. P. Sethna,
Europhys. Lett. {\bf 5}, 485 (1988).

\bibitem{Bray-89}
A. J. Bray,
J. Phys. A {\bf 22}, L81 (1989).

\bibitem{CMM-98}
F. Cesi, C. Maes, and F. Martinelli,
Comm. Math. Phys. {\bf 188}, 135 (1997); {\em ibid.}
 {\bf 189}, 323 (1997).

\bibitem{expG}
M. B. Salamon, P. Lin, and S. H. Chun,
Phys. Rev. Lett. {\bf 88}, 197203 (2002);
J. Deisenhofer, D. Braak, H.-A. Krug von Nidda, J. Hemberger, R. M. Eremina, 
V. A. Ivanshin, A. M. Balbashov, G. Jug, A. Loidl, T. Kimura, and Y. Tokura,
Phys. Rev. Lett. {\bf 95}, 257202 (2005);
P. Y. Chan, N. Goldenfeld, and M. Salamon,
Phys. Rev. Lett. {\bf 97}, 137201 (2006);
C. Magen, P. A. Algarabel, L. Morellon, J. P. Araujo, C. Ritter, M. R. Ibarra,
A. M. Pereira, and J. B. Sousa,
Phys. Rev. Lett. {\bf 96}, 167201 (2006); 
R.-F. Yang, Y. Sun, W. He, Q.-A. Li, and Z.-H. Cheng,
Appl. Phys. Lett. {\bf 90}, 032502 (2007);
W. Jiang, X. Zhou, G. Williams,  Y. Mukovskii, and K. Glazyrin,
Phys. Rev. Lett. {\bf 99}, 177203 (2007). 

\bibitem{MPV-02}
V. Mart{\'\i}n-Mayor, A. Pelissetto, and E. Vicari, 
Phys. Rev. E {\bf 66}, 026112 (2002).

\bibitem{FL-68}
M. E. Fisher and J. S. Langer,
Phys. Rev. Lett. {\bf 20}, 665 (1968).

\bibitem{CMPV-03-2}
P. Calabrese, V. Mart\'{\i}n-Mayor, A. Pelissetto, and E. Vicari,
Phys. Rev. E  {\bf 68}, 016110 (2003). 

\bibitem{footalpha}
Scaling corrections in randomly diluted $O(N)$-invariant spin models are
expected to vanish as $|T-T_c|^{-\alpha}$ for $N\ge 2$, since
the specific-heat critical exponent $\alpha$ is negative.
These corrections decay very slowly in the $XY$ and Heisenberg
cases, since
$\alpha=-0.0151(3)$ and $\alpha=-0.1336(15)$, respectively;
see: 
M. Campostrini, M. Hasenbusch, A. Pelissetto, and E. Vicari,
Phys.~Rev. B {\bf 74}, 144506 (2006); M. Campostrini, M. Hasenbusch,
A. Pelissetto, P. Rossi, and E. Vicari, Phys.~Rev. B {\bf 65} 144520
(2002); Ref.~\onlinecite{PV-02} for additional references.

\bibitem{Harris-74}
A.~B.~Harris, J. Phys. C {\bf 7}, 1671 (1974).


\bibitem{BFMMPR-99}
H.~G.~Ballesteros, L.~A.~Fern\'andez, V.~Mart\'{\i}n-Mayor, A.~Mu\~noz~Sudupe,
G.~Parisi, and J.~J.~Ruiz-Lorenzo,
J. Phys. A: Math. Gen. {\bf 32}, 1 (1999).


\bibitem{BFMMPR-98}
H. G. Ballesteros, L. A. Fern\'andez, V. Mart\'{\i}n-Mayor, A. Mu\~noz~Sudupe,
G. Parisi, and J. J. Ruiz-Lorenzo,
Phys. Rev. B {\bf 58}, 2740 (1998).

\bibitem{HPPV-07}
M. Hasenbusch, F. Parisen Toldin, A. Pelissetto, and E. Vicari,
J. Stat. Mech.: Theory Exp. P02016 (2007). 

\bibitem{HPV-07}
M. Hasenbusch, A. Pelissetto, and E. Vicari,
J. Stat. Mech.: Theor. Exp. P11009 (2007).

\bibitem{PV-00}
A. Pelissetto and E. Vicari,
Phys. Rev. B {\bf 62}, 6393  (2000).

\bibitem{BMMRY-87}
A. J. Bray, T. McCarthy, M. A. Moore, J. D. Reger, and A. P. Young,
Phys. Rev. B {\bf 36}, 2212 (1987).

\bibitem{McKane-94}
A. J. McKane, Phys. Rev. B {\bf 49}, 12003 (1994).

\bibitem{AMR-00}
G. \'Alvarez, V. Mart{\'\i}n-Mayor, and J. J. Ruiz-Lorenzo,
J. Phys. A {\bf 33}, 841 (2000).

\bibitem{PS-00} 
D.V. Pakhnin and A.I. Sokolov, Phys. Rev. B {\bf 61}, 15130 (2000).

\bibitem{CMPV-03}
P. Calabrese, V. Mart\'{\i}n-Mayor, A. Pelissetto, and E. Vicari,
Phys. Rev. E  {\bf 68}, 036136 (2003). 

\bibitem{CPV-eqst}
P. Calabrese, M. De Prato, A. Pelissetto, and E. Vicari,
Phys. Rev. B {\bf 68}, 134418 (2003).

\bibitem{CPV-cross}
P. Calabrese, P. Parruccini, A. Pelissetto, and E. Vicari,
Phys. Rev. E {\bf 69}, 036120 (2004).

\bibitem{BCBJ-04}
P. E. Berche, C. Chatelain, B. Berche, and W. Janke,
Eur. Phys. J. B {\bf 38}, 463 (2004).

\bibitem{FHY-00}
R. Folk, Yu. Holovatch, and T. Yavors'kii,
Phys. Rev. B {\bf 61}, 15144 (2000).

\bibitem{cpv-03}
P. Calabrese, A. Pelissetto, and E. Vicari,
Phys. Rev. B {\bf 68}, 092409 (2003).

\bibitem{EA-75}
S. F. Edwards and P. W. Anderson,
J. Phys. F {\bf 5}, 965 (1975).

\bibitem{HPPV-07-2}
M. Hasenbusch, F. Parisen Toldin, A. Pelissetto, and E. Vicari,
Phys. Rev. B {\bf 76}, 184202 (2007).

\bibitem{BAZ-74}
E. Br\'ezin, D. J. Amit, and J. Zinn-Justin,
Phys. Rev. Lett. {\bf 32}, 151 (1974).

\bibitem{BLZ-74}
E. Br\'ezin, J. C. Le Guillou, and J. Zinn-Justin,
Phys. Rev. Lett. {\bf 32}, 473 (1974).

\bibitem{CPRV-98}
M. Campostrini, A. Pelissetto, P. Rossi, and E. Vicari,
Phys. Rev. E {\bf 57}, 184 (1998).

\bibitem{FS-75}
R. A. Ferrell and D. J. Scalapino,
Phys. Rev. Lett. {\bf 34}, 200 (1975).

\bibitem{Bray-76}
A. J. Bray, Phys. Rev. B {\bf 14}, 1248 (1976).

\bibitem{FB-67}
M. E. Fisher and R. J. Burford,
Phys. Rev. {\bf 156}, 583 (1967).

\bibitem{TF-75}
H. B. Tarko and M. E. Fisher,
Phys. Rev. Lett. {\bf 31}, 926 (1973);
Phys. Rev. B {\bf 11}, 1217 (1975).

\bibitem{FB-79}
R. A. Ferrell and J. K. Bhattacharjee,
Phys. Rev. Lett. {\bf 42}, 1505 (1979).

\bibitem{BBC-82}
D. Beysens, A. Bourgou, and P. Calmettes,
Phys. Rev. A {\bf 26}, 3589 (1982).

\bibitem{Abe-68}
R. Abe, Progr. Theor. Phys. {\bf 39}, 947 (1968).

\bibitem{GL-76}
G.~Grinstein and A.~Luther,
Phys. Rev. B  {\bf 13}, 1329  (1976).

\bibitem{DeDominicis-78}
C. De Dominicis, Phys. Rev. B {\bf 18}, 4913 (1978).

\bibitem{FM-06}
R. Folk and G. Moser,
J. Phys. A: Math. Gen. {\bf 39}, R207 (2006).

\bibitem{CG-05}
P. Calabrese and A. Gambassi,
J. Phys. A {\bf 38}, R133 (2005).

\bibitem{pertcalc}
G. Grinstein, S. Ma, and G. F. Mazenko, Phys. Rev. B {\bf 15}, 258 (1977); 
U. Krey, Z. Phys. B {\bf 26}, 355 (1977).
V. V. Prudnikov, J. Phys. C {\bf 16}, 3685 (1983); I. D. Lawrie and V. V.
Prudnikov, J. Phys. C {\bf 17}, 1655 (1984);
K. Oerding and H. K. Janssen, J. Phys. A {\bf 28}, 4271 (1995); 
H. K. Janssen, K. Oerding, and E. Sengespeick, 
J. Phys. A {\bf 28}, 6073 (1995);
V. V. Prudnikov, S. V. Belim, E. V. Osintsev, and A. A. Fedorenko,
Fiz. Tverd. Tela {\bf 40}, 1526 (1998) 
[English translation, Phys. Solid State {\bf 40}, 1383 (1998)];
P. Calabrese and A. Gambassi Phys. Rev. B {\bf 66}, 212407 (2002);
G. Schehr and R. Paul, Phys. Rev. E {\bf 72}, 016105 (2005).

\end{thebibliography}
\end{document}